\documentclass[aps]{revtex4}
\usepackage{amssymb}
\usepackage{graphicx}
\usepackage{bm}
\usepackage{amsmath}

\begin{document}
\title{General Nonlocality in Quantum Fields}
\author{Hai-Jun Wang}
\address{Center for Theoretical Physics and School of Physics, Jilin University,\\
Changchun 130023, China}

\begin{abstract}
The natural recognition of quantum nonlocality follows from the fact that a
quantum wave is spatially extended. The waves of fermions display
nonlocality in low energy limit of quantum fields. In this {\it ab initio}
paper we propose a complex-geometry model that reveals the affection of
nonlocality on the interaction between material particles of spin-$1/2$. To
make nonlocal properties appropriately involved in a quantum theory, the
special unitary group $SU(n)$ and spinor representation $D^{(1/2,1/2)}$ of
Lorentz group are generalized by making complex spaces---which are spanned
by wave functions of quantum particles---curved. The curved spaces are
described by the geometry used in General Relativity by replacing the real
space with complex space and additionally imposing the analytic condition on
the space. The field equations for fermions and for bosons are respectively
associated with geodesic motion equations and with local curvature of the
considered space. The equation for fermions can restore all the terms of
quadratic form of Dirac equation. According to the field equation it is
found that, for the U(1) field [generalized Quantum Electrodynamics (QED)],
when the electromagnetic fields $\vec E$ and $\vec B$ satisfy $\vec E^2-\vec
B^2\neq 0$, the bosons will gain masses. In this model, a physical region is
empirically defined, which can be characterized by a determinant occurring
in boson field equation. Applying the field equation to $U(3)$ field
[generalized Quantum Chromodynamics (QCD)], the quark-confining property can
be understood by carrying out the boundary of physical region. And it is
also found out that under the conventional form of interaction vertex, $%
\gamma _\mu A^\mu $, only when the colour group $SU(3)$ is generalized to $%
U(3)$ is it possible to understand the strongly bound states of quarks.

\end{abstract}
\maketitle

\section{Introduction}

Nonlocality is an important phenomenon in nature, particularly in quantum
world. The direct recognition of quantum nonlocality comes from the fact
that a quantum wave is spatially extended, in contrast to the point--model
for classical particles. In this paper we mainly discuss how the nonlocality
affects the interactions between material particles of spin-$1/2$. The
problem is intriguing since the nonlocality has been gleamingly implied by
the renormalization of conventional Quantum Field Theory (CQFT), whence most
relevant calculations have to be regulated by momentum cutoff to contain the
non-point effect. The technique however, is usually available only at
high--energy scale, the case where the wavelengths of particles are ultra
short. Here we take into account the nonlocal effect emerging within the
range of interactions---possibly a few wavelengths; but we don't get
involved in the hotly discussed long--distance effects relating to entangled
states and their applications---such as quantum information, quantum
communication and quantum computation etc..

Up to date, we have recognized that one cannot accurately measure the
spatial coordinates of a proton by making an accelerated electron as probe,
unless its wavelength is much shorter than the ''diameter'' of the proton.
But the proton would be smashed and some other hadrons will be involved in
the final state (and thus the scattering becomes inelastic) if making the
electron's wavelength short enough. In the case of elastic scattering, the
detected proton becomes a {\bf singularity} for the electron's wave. The
reason may be that, in the measurements, the quantity (coordinates) we
inquire is not at the same spatial level as that the quantum entities
settled in---the coordinate is a four-dimension quantity but the electron's
or proton's wave is eight-dimension, or put it in mathematical terminology,
the quantity we inquire is real but a quantum object is complex.

It is concluded from purely mathematical point of view that, only located in
a space with dimension equal to or larger than that of the detected objects
can an observer get complete information of direct measurement. As a
tentative method and a starting point, in this paper we propose an {\bf %
equal observer}, e.g. an electron, is also put into the Hilbert space to
observe another electron or other fermions such as protons. Presumably, they
are at the same spatial level. Therefore the electron can use the metric
(gauge) appropriate for the observed objects to measure physical
observables. The method of {\bf equal observer} is conducive to describing
the observed quantum wave ({\bf nonlocal entity}) as a whole with possibly
less interaction-information lost, unlike in conventional quantum mechanics
(CQM) where quantum wave is expressed on the basis of space-time points. The
dynamics for the equal observer of a quantum wave is believed to be
different from CQM.

In this paper we employ the similarity between quantum {\bf singularity} and
gravitational {\bf singularity} to describe how one fermion observes
(interacts with) another fermion, and dynamically apply the formalism of
General Relativity (GR) by generalizing its space from real to complex [Fig.
1]. As for the elastic scattering of electron and proton, in calculating the
radiative corrections to the lowest order of scattering process by employing
Quantum Electrodynamics (QED), we encounter the divergence rooted from
leading-order potential form $\frac 1r$ while making momentum $\mid \vec p%
\mid \rightarrow \infty $. In calculating the collision of two heavy
celestial bodies by using GR, the similar singularity rooted also from the
form $\frac 1r$ is encountered, but there the puzzle of divergence is
automatically circumvented by carrying out a horizon, the outer of which is
physical region, and the inner of which, now known as black hole region, is
unphysical. Quantum mechanically, the nonlocal region is usually
unobservable in our space-time, and thus unphysical. Enlightened by such
physical scenario, we expect to define physical region for elemental
fermions in complex space.

In analogy to GR, the principle of nonlocality for two interacting
fermions
is: {\bf There always exists a complex frame for observer }(one fermion){\bf %
\ in which the observed fermion }(another fermion){\bf \ looks like a plane
wave, no matter the existence of interaction.}

CQFT itself can also lead us to perceive the implicit and profound
relationship between nonlocality (quantum wave) and complex-curvature.
Generally, we interpret the scattering matrix between initial state $\mid
i\rangle $ and final state $\mid f\rangle $ as $S_{fi}=\langle f\mid \infty
\rangle =\langle f\mid S\mid i\rangle $, where $\mid i\rangle =$ $\mid
-\infty \rangle $ can be any state of a complete set. In this formalism, the
operator $S$ (or alternatively, the Hamiltonian) is assumed known. Then the
matrix elements $S_{fi}$--whose square is proportional to transition rate
between initial and final states--can be evaluated. Whereas from an
equal--observer angle, all the states $\mid i\rangle $ are known and the
state $\mid \infty \rangle _f$ can be observed, so the operator $S$ can be
carried out formally $S_f=%
\mathop{\textstyle \sum }
\nolimits_i\mid \infty \rangle _f\langle i\mid $, consequently the
interaction becomes known. This latter opposite thought reminds us of the
physics in GR, where the force can be realized by the curving of space-time.
So, if only the $S$--matrix is defined locally in complex-space (a quantum
wave is viewed as a point in such space, and as a whole), the differential
geometry for {\bf nonlocal entity} would definitely occur. [Note: For
convenience, in what follows we will not employ the language of $S$--matrix,
though relevant.] The further understanding of the relationship between
nonlocality and curvature is achieved in section 10, where the local
conservation laws don't exist. In summary, one will gradually be aware of
that all of the above intuitive knowledge on nonlocality could be {\it %
transferred} to and {\it interpreted} by the Complex Geometry we used, {\it %
via} the motion equation for fermions and the field equation for
bosons in section 8$\sim$9.

The main results of the paper are as follows: From our dynamical motion
equation we can restore all the terms appearing in the quadratic form of
Dirac equation. For the $U(1)$ field [generalized QED], if electromagnetic
fields $\vec B$ and $\vec E$ satisfy $\vec E^2-\vec B^2\neq 0$, the bosons
will gain masses. Based on the discussion of physical region, we can attain
two qualitative understandings of quark confinement.

In order not to make readers confused by mathematical details, we only list
some main results of the real and complex geometry with detailed
explanations and the necessary calculating techniques, then immediately
apply them to physics problems---discussion of dynamical equations and the
definition of physical region. The remainder of this paper is arranged as
follows. In sections 2-5 we introduce the necessary calculating techniques
of geometry, the readers familiar with the contents can neglect this part
and directly turn to section 6. Then in sections 6-9 we apply them to
construct the motion equations for fermions and field equation for bosons.
The conservation laws are sketched in section 10. In section 11, the
physical region is defined for $U(1)$ field [generalized QED] and $U(3)$
field [generalized QCD]. Consequently the qualitative understandings to
confinement of quarks is presented under the approximation of interaction
vertex $\gamma _\mu A^\mu $. Finally a concluding section is presented to
summarize the paper and give some remarks on the applicability of the
theory. In appendix A, the derivation of motion equation for fermions is
elaborated.

\section{A Useful Tool: Exterior Differentiation $d$}

Exterior product proves to be a powerful tool in calculating tensors. It
obeys the rules of Grassman algebra. One of its usages is to give rise to
new tensors of higher order. For example, let $\{x_1,x_2,\cdots ,x_n\}$ be
of coordinates for a space, and $f(x_1,x_2,\cdots ,x_n)$ an arbitrary scalar
function with respect to these coordinates, then the total differential of
the function $f$ is d$f=\sum_i\frac{\partial f}{\partial x_i}%
dx_i,i=1,2,\cdots ,n$. We call $(\frac{\partial f}{\partial x_1},\frac{%
\partial f}{\partial x_2},\cdots ,\frac{\partial f}{\partial x_n})_p$ a
one-order tensor, or a vector, and $(dx_1,dx_2,\cdots ,dx_n)_p$ the basis in
the neighborhood of a given point $p$. In terms of exterior product the d$f$
is called differential 1-form. Now we can construct 2-form on the basis of
the 1-form: dd$f=\sum_{ij}\frac{\partial \partial f}{\partial x_i\partial x_j%
}dx_i\wedge dx_j,i=1,2,\cdots ,n$. The sign $\wedge $ denotes the exterior
product. It satisfies the antisymmetric rule: $dx_i\wedge dx_j=-dx_j\wedge
dx_i$, according to which one can easily conclude dd$f=0$. Conventionally,
an $n$-form is written as
\begin{equation}
T_{ij\cdots k}dx^i\wedge dx^j\wedge \cdots \wedge dx^k\text{.}
\tag{2.1}
\end{equation}
The repeated index henceforth in this whole paper means summation. Here the $%
T_{ij\cdots k}$ are components of the $k-$order tensor, and the exterior
products $dx^i\wedge dx^j\wedge \cdots \wedge dx^k$ are the bases.

The above forms can be readily extended to complex variables, e.g., the two
complex variables \{{\em z,\thinspace \=z}\}. And the only 2-form of these
complex variables is $g(z,\bar z)$ \text{d}$z$ $\wedge ${\em \thinspace }%
\text{d}$\bar z$ where $g(z,\bar z)$ is a function of {\em z} and {\em %
\=z.} Here {\em z} and {\em \=z} are independent variables.

\section{Cauchy-Riemann Condition for Complex
Manifold}

Usually a space is called $n$-dimension manifold if it can be formed as
smoothly as possible by affixing many infinitesimally flat patches of ${\bf R%
}^n$($n$-dimension real space), ${\bf C}^n$($n$-dimension complex space), or
other sorts of $n$-dimension spaces. In fact in this paper we are only
concerned about a particular sort of differential manifolds, but for
convenience we will call it manifold without specification.

Above all, an important property for a given space is the
existence of derivatives, which is essential to build up a
description of differential geometry. As for Hilbert space, the
existence of derivatives may be determined by that of wave
function $\psi(x)$ in it. In general, the existence of derivatives
of a real function can be understood intuitively. However, for a
complex function, that demands the Cauchy-Riemann condition be
satisfied. When the complex derivative can be defined
''everywhere,'' the function is said to be {\bf analytic}. Now we
review it in respect of complex geometry.

Conventionally, we identify an $n$-dimension complex manifold with a $2n$%
-dimension real manifold. The complex coordinates and real coordinates have
the relation
\begin{equation}
{z}^\alpha =x^\alpha +iy^\alpha ,{\bar z}^\alpha =x^\alpha
-iy^\alpha ,\;\alpha =1,\cdots ,n \text{ ,} \tag{3.1}
\end{equation}
then the following 1-form is written straightforwardly

\begin{equation}
\text{d}{z}^\alpha =\text{d}x^\alpha +i\text{d}y^\alpha
,\;\text{d}{\bar z}^\alpha =\text{d}x^\alpha -i\text{d}y^\alpha
\text{ ,} \tag{3.2}
\end{equation}
and subsequently
\begin{equation}
\frac \partial {\partial x^\alpha } =\frac{\partial {z}^\alpha }{%
\partial x^\alpha }\frac \partial {\partial {z}^\alpha }+\frac{\partial
{\bar z}^\alpha }{\partial x^\alpha }\frac \partial {\partial {\bar z%
}^\alpha }=\frac \partial {\partial {z}^\alpha }+\frac \partial {%
\partial {\bar z}^\alpha }\text{ ,}   \tag{3.3 a}
\end{equation}
\begin{equation}
\frac \partial {\partial y^\alpha }=\frac{\partial {z}^\alpha }{%
\partial y^\alpha }\frac \partial {\partial {z}^\alpha }+\frac{\partial
{\bar z}^\alpha }{\partial y^\alpha }\frac \partial {\partial {\bar z%
}^\alpha }=i\frac \partial {\partial {z}^\alpha }-i\frac \partial {%
\partial {\bar z}^\alpha }\text{ .} \tag{3.3 b}
\end{equation}
The reverse of the above equations yields
\begin{equation}
\frac \partial {\partial { z}^\alpha }\stackrel{d}{\equiv
}\partial
_\alpha =\frac 12(\frac \partial {\partial x^\alpha }-i\frac \partial {%
\partial y^\alpha })\text{ ,}  \tag{3.4 a}
\end{equation}
\begin{equation}
\frac \partial {\partial { \bar z}^\alpha }\stackrel{d}{\equiv
}\bar
\partial _\alpha =\frac 12(\frac \partial {\partial x^\alpha }+i\frac
\partial {\partial y^\alpha })\text{ ,}  \tag{3.4 b}
\end{equation}
here the sign $d$ above the equals sign means a {\bf definition}.
Perform exterior-product upon an arbitrary complex scalar function
$f({ z}^\alpha ,{ \bar z}^\alpha )$, it results in a 1-form
\begin{equation}
\text{d}f=\partial _\alpha f\,\text{d}{ z}^\alpha +\bar
\partial _\alpha f\,\text{d}{ \bar z}^\alpha
\stackrel{d}{\equiv }\partial f+\bar \partial f\text{ .}
\tag{3.5}
\end{equation}
Make $f=u+i\upsilon $, we note that $\bar \partial f=0$ is just the
Cauchy-Riemann relation
\begin{equation}
\frac{\partial { u}}{\partial x^\alpha }=\frac{\partial { \upsilon }}{%
\partial y^\alpha },\;\frac{\partial { u}}{\partial y^\alpha }=-\frac{%
\partial { \upsilon }}{\partial x^\alpha }  \tag{3.6}
\end{equation}

Another important property of a given space is its transformation.
Commonly, the vector fields spanned by basis $(dx_1,dx_2,\cdots
,dx_n)$ [real space] or by basis \{$($d${ z}^\alpha ,$d${ \bar
z}^\alpha ),$ $\alpha =1,\cdots ,n$\} [complex space] on manifolds
are the objects responsible for addressing the geometry of the
manifolds. How do these fields transform from one point to another
infinitesimal neighboring point of the manifolds? The
transformations are usually constrained by some groups as done in
the well-known classical mechanics or quantum fields. We will
return to this topic later. Now let's turn to the affection of
Cauchy-Riemann relation on a vector basis. Ignoring some
unnecessary details, we assume that the transformation of the
$n$-dimension complex manifold has the general form
\begin{equation}
{ z}^\alpha \rightarrow { z}^{\prime \alpha }=f^\alpha ({ z}^1,{
\bar z}^1;{ z}^2{\frak ,}{ \bar z}^2;\cdots ;{ z}^n,{ \bar z}^n)%
\text{ .}  \tag{3.7}
\end{equation}
As an application of Eq.(3.6), replacing the function $f$ by $f^\alpha $,
and writing the $f^\alpha $ of Eq.(3.7) explicitly as $f^\alpha =u^\alpha
+i\upsilon ^\alpha $, the form of the Eq.(3.6) now reads
\begin{equation}
\frac{\partial { u}^\alpha }{\partial x^\beta }=\frac{\partial { %
\upsilon }^\alpha }{\partial y^\beta },\;\frac{\partial { u}^\alpha }{%
\partial y^\beta }=-\frac{\partial { \upsilon }^\alpha }{\partial x^\beta
},\;\alpha ,\beta =1,\cdots ,n\text{ .}  \tag{3.8}
\end{equation}
Make $y^\alpha =x^{n+\alpha },$ ${ \upsilon }^\alpha ={
u}^{n+\alpha }$
and introduce a $2n\times 2n$ matrix $J$%
\begin{equation}
J=\left(
\begin{array}{cc}
0 & -I_n \\
I_n & 0
\end{array}
\right) \text{ ,}  \tag{3.9}
\end{equation}
which satisfies $J^2=-I_{2n}$, $I_{2n}$ denoting the identity matrix of rank
$2n$, then the Cauchy-Riemann relation (3.8) can be written
\begin{equation}
J_k^j\frac{\partial { u}^k}{\partial x^l}=\frac{\partial { u}^j}{%
\partial x^k}J_l^k,\;j,k,l=1,\cdots ,2n\text{ ,}  \tag{3.10}
\end{equation}
and correspondingly an identical relation holds
\begin{equation}
J_l^j=\frac{\partial { u}^j}{\partial x^k}J_n^k\frac{\partial x^n}{%
\partial { u}^l}\text{ .}  \tag{3.11}
\end{equation}
Write out the Eq. (3.10) alternatively as
\begin{equation}
J_l^k\frac{\partial { u}^j}{\partial x^k}=\frac{\partial (J_k^j{ u}^k)%
}{\partial x^l}\text{ ,}  \tag{3.12}
\end{equation}
and regardless of $J$ on ${ u}^k$, we obtain

\begin{equation}
J\frac \partial {\partial x^\alpha }=\frac \partial {\partial y^\alpha },\,J%
\frac \partial {\partial y^\alpha }=-\frac \partial {\partial x^\alpha }%
\text{ .}  \tag{3.13}
\end{equation}
Combining it with the equation (3.4) gives

\begin{equation}
J\frac \partial {\partial { z}^\alpha } =-i\frac \partial
{\partial { z}^\alpha }\text{ ,}  \tag{3.14 a}
\end{equation}
\begin{equation}
J\frac \partial {\partial { \bar z}^\alpha } =i\frac \partial
{\partial { \bar z}^\alpha }\text{ .}  \tag{3.14 b}
\end{equation}
Consistently the results (3.13)\symbol{126}(3.14) can also be
expressed
using the basis $(dx_1,\cdots ,dx_n;dy_1,\cdots ,dy_n)$ and $\{$d${ z}%
^\alpha $,d${ \bar z}^\alpha \}$, respectively as
\begin{equation}
J\text{d}x^\alpha =-\text{d}y^\alpha ,J\text{d}y^\alpha
=\text{d}x^\alpha \text{ ,}  \tag{3.15}
\end{equation}
and
\begin{equation}
J\text{d}{ z}^\alpha =-\text{d}y^\alpha +i\text{d}x^\alpha =i\text{d}%
{ z}^\alpha \text{ ,}  \tag{3.16 a}
\end{equation}
\begin{equation}
J\text{d}{ \bar z}^\alpha =-\text{d}y^\alpha -i\text{d}x^\alpha =-i%
\text{d}{ \bar z}^\alpha \text{.}  \tag{3.16 b}
\end{equation}
From Eq. (3.16), we note that the d${ z}^\alpha $ and d${ \bar z}%
^\alpha $ are eigen forms of the matrix $J$, and the relations are
independent of the choice of the coordinates. In this respect $J$ is called
the {\bf complex structure} of the complex manifold.

A manifold with complex structure is called {\bf almost} complex manifold.

The above mathematical formulae will become more meaningful if we replace $%
{ z}^\alpha $ and ${ \bar z}^\alpha $ by the wave function ${ \psi }%
^\alpha $ and ${ \bar \psi }^\alpha $ from the Hilbert space
(Manifold), which will be recognized in following sections. For
convenience, henceforth we will put ${ \psi }^\alpha $ in a more
general space, with Hilbert space only as a special case.

\section{A Route to Construct Differential Geometry for a Space}

Before giving the general definition of metric (gauge), it is helpful to
recall its definition in three-dimensional (3-D) space. Using the Cartan
Method of Moving Frames [1], let every point $\vec x=(x_1,x_2,x_3)$ in 3-D
space correspond to a frame $\{\vec e_1,\vec e_2,\vec e_3\}$as following. If
the point $\vec x$ varies with a curvilinear net defined by $(u_1,u_2,u_3),$
i.e., $x_i=x_i(u_1,u_2,u_3)$, the total differential of $\vec x$ then takes
the form
\begin{equation}
d\vec x=\frac{\partial \vec x}{\partial u_i}du_i,\;i=1,2,3\text{
,} \tag{4.1}
\end{equation}
if setting $\frac{\partial \vec x}{\partial u_i}=P^{ij}\vec e_j$ and $\omega
^j=du_iP^{ji}$, then from Eq.(4.1) the point $\vec x$ relates to frame $\{%
\vec e_1,\vec e_2,\vec e_3\}$ as follows
\begin{equation}
d\vec x=\omega ^i\vec e_i\text{ .}  \tag{4.2}
\end{equation}

Now define metric tensor as
\begin{equation}
g_{ij}=(\vec e_i,\vec e_j)\text{ ,}  \tag{4.3}
\end{equation}
the parenthesis denotes the inner product of common sense. Combining these
definitions, the metric of $d\vec x$ reads
\begin{equation}
ds^2=(d\vec x,d\vec x)=(\omega ^i\vec e_i,\omega ^j\vec
e_j)=g_{ij}\omega ^i\omega ^j\text{.}  \tag{4.4}
\end{equation}
From Eq.(4.2), the differential of basis $\vec e_i$ can be formally written
as
\begin{equation}
d\vec e_i=\omega _i^{\;j}\vec e_j\text{ ,}  \tag{4.5}
\end{equation}
where $\omega _i^j$ is also a 1-form similar to $\omega ^i$, then the
differential for $g_{ij}$ can be calculated as follows
\begin{equation}
dg_{ij}=(d\vec e_i,\vec e_j)+(\vec e_i,d\vec e_j)=g_{kj}\omega
_i^{\;k}+g_{ik}\omega _j^{\;k}\text{ .}  \tag{4.6}
\end{equation}
The above general forms is suitable for any frame $\{\vec e_1,\vec e_2,\vec e%
_3\}$. Now let's turn to some special cases relevant to concrete frames.

A well-known metric form for coordinates frames is the orthogonal one, i.e. $%
g_{ij}=(\vec e_i,\vec e_j)=\delta _{ij}$. With this relation, Eq.(4.4)
becomes
\begin{equation}
ds^2=\omega ^i\omega _i\text{ ,}  \tag{4.7}
\end{equation}
and Eqs. (4.6), (4.7) give rise to
\begin{equation}
\omega _i^{\;k}=-\omega _k^{\;i}\text{ .}  \tag{4.8}
\end{equation}
Another general/important frame is the {\bf natural frame}, in which $\vec e%
_i$ is defined by
\begin{equation}
\vec e_i=\frac{\partial \vec x}{\partial u^i}\text{,}  \tag{4.9}
\end{equation}
then (4.1) is directly written as
\begin{equation}
d\vec x=du^i\vec e_i\text{,}  \tag{4.10}
\end{equation}
and (4.4) reads
\begin{equation}
ds^2=g_{ij}du^idu^j\text{ .}  \tag{4.11}
\end{equation}
The natural frame is of elicitation in studying manifold, which
can help us extend the above discussion to n-dimensional space.
The space continuously generated by curvilinear net
$\{u^1,u^2,\cdots ,u^n\}$ is a manifold. Now let's turn to its
natural bases $\{\vec e_i=\frac{\partial \vec x}{\partial u^i}\}$.

According to the terminology of quantum mechanics, $\vec e_i$ is made by two
parts: a matrix (operator) $\frac \partial {\partial u^i}$ and a vector $%
\vec x$. Since the operator parts $\frac \partial {\partial u^i}$ are the
same for different points $\vec x$, we usually write $\vec e_i=\frac \partial
{\partial u^i}$ for general cases while $\vec e_i=\frac \partial {\partial
u^i}\mid _{\vec x}(=\frac{\partial \vec x}{\partial u^i})$ for a given point
$\vec x$. In this sense, the general form of inner product $g_{ij}$ becomes
\begin{equation}
g_{ij}=(\frac \partial {\partial u_i},\frac \partial {\partial
u_j})\text{ .} \tag{4.12}
\end{equation}
For a given point $p$ on manifold, $g_{ij}(p)$ gives rise to a mapping that
maps two arguments $\frac \partial {\partial u_i}\mid _p$ and $\frac \partial
{\partial u_j}\mid _p$ to a real or complex number.

Then what is the relationship between the basis $du^i$--- which has been
mentioned in section II---and this $\frac \partial {\partial u_i}$? That can
be recognized by the integral $\int du^i\frac \partial {\partial u^j}%
f(u)=f(u)$ where $f$ is an arbitrary function, the integrand $f(u)$ is
unaltered after integration. So it is reasonable to write
\begin{equation}
(du^i,\frac \partial {\partial u^j})=\delta _{\;j}^i\text{ .}
\tag{4.13}
\end{equation}
The contravariant metric tensor $g^{ij}$ is thus defined by
\begin{equation}
g^{ij}=(du^i,du^j)\text{ ,}  \tag{4.14}
\end{equation}
with $g^{ij}g_{jk}=\delta _{\;k}^i$. The above arguments equally hold in
complex space.

For a given space, there may exist several ways to define its metric form,
{\it i.e.,} the gauge to measure the space. In the physical respect, we are
only interested in the forms like Eq. (4.14), even for complex manifold.

The construction of metrics includes another important feature, i.e. the
constructed metric $ds^2$ should be of geometrically invariant quantity with
respect to the transformation between frames of the different points of the
manifold/space. For a given space, there may exist many ways to jump from
one point to another point in its infinitesimal neighborhood. To address the
manner of the jump is to determine the transformation group between the
corresponding frames. The transformation group is also a key element to
define manifold, different groups correspond to different manifolds.

Now let's turn to the general form of differential geometry. Perform the
infinitesimal group transformation upon the vector basis $\frac \partial {%
\partial x_i}=e_i$, the change of $e_i$ is then analogous to Eq. (4.5)
\begin{equation}
de_j=\Gamma _{\;j}^i\,e_i\text{ ,}  \tag{4.15}
\end{equation}
$\Gamma _{\;\alpha }^\beta $ is 1-form (the similar meaning as the above $%
\omega _i^{\;j}$) on the manifold. Conventionally, we take the explicit form
$\Gamma _{\;i}^j=\Gamma _{ki}^jdx^k,$ in which $\Gamma _{ik}^j$ (not a
tensor) is known as {\bf connection}, referring to the manner how to affix
the infinitesimal flat-space-patches to form a larger curved space.
According to above definition Eq. (4.15), the variation of a vector $X=\xi
^ie_i$ under the transformation yields
\begin{equation}
dX=d\xi ^ie_i+\xi ^ide_i=(d\xi ^i+\xi ^j\Gamma _{\;j}^i\,)e_i=(D\xi ^i)e_i%
\text{ ,}  \tag{4.16}
\end{equation}
where $D$ is known as covariant differential. The above
expressions indicate that the notations $de_\alpha $, $De_\alpha $
and $\Gamma _{\;\alpha }^\beta $ are identical:$De_\alpha =\Gamma
_{\;\alpha }^\beta \,e_\beta $, which leads us to a conventional
denotation $D=\Gamma _{\;\alpha }^\beta \,$ where $D$ and $\Gamma
_{\;\alpha }^\beta $ are both viewed as operators. If for all
components $D\xi ^i=0$ is met, the differential $dX$ is called the
parallel displacement for vector $X$, and the path along which the
displacement would occur is solvable from the equation $D\xi
^i=0$.

The connection $\Gamma _{\;\alpha }^\beta $ can be carried out by demanding
that it should preserve the inner product of two vector $X$ and $Y$
\begin{equation}
d(X,Y)=0\text{ ,}  \tag{4.17}
\end{equation}
if they are parallelly displaced, i.e. $dX=dY=0$. In the above equation
\begin{equation}
(X,Y)=(\xi ^ie_i,\eta ^je_j)=g_{ij}\xi ^i\eta ^j\text{ ,}
\tag{4.18}
\end{equation}
and $\xi ^i$ and $\eta ^j$ are 1-form. We see that
\begin{equation}
0=d(X,Y)=dg_{ij}\xi ^i\eta ^j+g_{ij}d\xi ^i\eta ^j+g_{ij}\xi
^id\eta ^j \tag{4.19}
\end{equation}
leads to
\begin{equation}
(dg_{ij}-g_{kj}\Gamma _{\;i}^k-g_{ik}\Gamma _{\;j}^k)\xi ^i\eta
^j=0\text{ ,} \tag{4.20}
\end{equation}
i.e.
\begin{equation}
Dg_{ij}=dg_{ij}-g_{kj}\Gamma _{\;i}^k-g_{ik}\Gamma
_{\;j}^k=0\text{.} \tag{4.21}
\end{equation}
The parallel displacement $D\xi ^i=0$ and the conclusion that the
connection preserves the metric $(X,Y)$, i.e. Eq. (4.21), are
equivalent---they can bring out each other. Note that all the
components of $\Gamma _{\;j}^i$ form a matrix $\Gamma =(\Gamma
_{\;j}^i)$, and likewise components of $g_{ij}$ form a matrix $G$,
Eq. (4.21) can be written in terms of matrices
\begin{equation}
dG=\Gamma ^t\cdot G+G\cdot \Gamma \text{ ,}  \tag{4.22}
\end{equation}
$\Gamma ^t$ means the transpose of the matrix $\Gamma $. The relation Eq.
(4.22) puts a way--which we will elucidate later--to carry out the
connection $\Gamma _{jk}^i$ with respect to the metric $g_{ij}$.

Let's turn to the most important quantity of manifold--the curvature
tensor--which is an invariant 2-form under a certain transformation group
that specifies the manifold. Conventionally, the curvature matrix is defined
by
\begin{equation}
\Omega =(\Omega _\alpha ^\beta )=d\Gamma -\Gamma \wedge \Gamma
\text{ .} \tag{4.23}
\end{equation}
Taking $\Omega _\alpha ^\beta =R_{\alpha jk}^\beta dx^j\wedge dx^k$ and
bearing in mind that $\Gamma _{\;\alpha }^\beta =\Gamma _{k\alpha }^\beta
dx^k$, we can express it explicitly
\begin{equation}
R_{\alpha jk}^\beta =\Gamma _{k\alpha ,j}^\beta -\Gamma _{j\alpha
,k}^\beta +\Gamma _{j\gamma }^\beta \Gamma _{k\alpha }^\gamma
+\Gamma _{k\gamma }^\beta \Gamma _{j\alpha }^\gamma \text{ ,}
\tag{4.24}
\end{equation}
the denotation $,j$ in subscripts stands for the derivative with respect to
the $j$-th variable. Then by Tr$\Omega ,$ i.e. making $\alpha =\beta $ in
(4.24) [summation convention is applied too.] one gets the Ricci tensor
\begin{equation}
R_{jk}=R_{\alpha jk}^\alpha \text{\ .}  \tag{4.25}
\end{equation}
Above formulae pertain to all manifolds with the metric form like Eq. (4.14).

\section{Three useful complex manifolds}

\subsection{Connections for almost complex manifold}

The analytic property of complex manifold is determined completely by the
complex structure, so any connection possessed by complex manifold is
demanded to preserve the complex structure. i.e.
\begin{equation}
J\,D_\alpha =D_\alpha \,J\text{ ,}  \tag{5.1}
\end{equation}
where $D_\alpha $ means performing differential on a given basis $e_\alpha $%
, i.e. $D_\alpha \,e_\beta =\Gamma _{\alpha \beta }^\gamma e_\gamma $. Apply
Eq. (5.1) to Eq. (4.15), remembering the Eq.(3.14) [Please note: for a
complex manifold, we use denotations $e_\alpha =\frac \partial {\partial
{ z}^\alpha }$ and $e_{\bar \alpha }=\frac \partial {\partial { \bar z}%
^\alpha }$ hereafter], then
\begin{equation}
J\,D_\alpha \,e_\beta =J\Gamma _{\alpha \beta }^\gamma e_\gamma
=\Gamma _{\alpha \beta }^\gamma Je_\gamma =-i\Gamma _{\alpha \beta
}^\gamma e_\gamma =D_\alpha \,Je_\beta \text{.}  \tag{5.2}
\end{equation}
However, if we apply Eq.(5.1) to $e_{\bar \alpha }$, the result becomes
\begin{equation}
J\,D_\alpha \,e_{\bar \beta }=J\Gamma _{\alpha \bar \beta }^\gamma
e_\gamma =\Gamma _{\alpha \bar \beta }^\gamma Je_\gamma =-i\Gamma
_{\alpha \bar \beta }^\gamma e_\gamma =-D_\alpha \,Je_{\bar \beta
}\text{ ,}  \tag{5.3}
\end{equation}
which violates Eq.(5.1). Therefore, Eq.(5.1) holds only if
\begin{equation}
\Gamma _{\alpha \bar \beta }^\gamma \equiv 0\text{ .}  \tag{5.4 a}
\end{equation}
Replace $\Gamma _{\cdot \;\cdot }^\gamma $ by $\Gamma _{\cdot \;\cdot }^{%
\bar \gamma }$ and repeat the same procedure, we have
\begin{equation}
\Gamma _{\alpha \beta }^{\bar \gamma }\equiv 0\text{ .}  \tag{5.4
b}
\end{equation}

Eqs.(5.4) suggest only the 1-form $\Gamma _{\;\alpha }^\beta $ and
$\Gamma _{\;\bar \alpha }^{\bar \beta }$ are permitted in the
complex manifold. If the connection of a manifold preserves the
complex structure $J$, then the manifold is called almost complex
manifold, this statement is equivalent to
aforementioned definition. Furthermore if the repeat of the differential $%
\partial $, i.e. $\partial ^2$ always turns to null, $\partial ^2=0$ (other
equivalent conditions are $\bar \partial ^2=0$ and $\partial \bar \partial =-%
\bar \partial \partial $), then the almost complex manifold becomes complex
manifold. (The proof of these statements is omitted in this paper.) In this
paper only complex manifolds are concerned.

$\Gamma _{\;\alpha }^\beta =\Gamma _{\gamma \alpha }^\beta d{
z}^\gamma $
is called $[1,0]$ type connection, and $\Gamma _{\;\alpha }^\beta =\Gamma _{%
\bar \gamma \alpha }^\beta d{ \bar z}^\gamma $ is called $[0,1]$
type connection. The group transformation cannot change the type
of connection.
In this paper, for simplicity we employ only $[1,0]$ type connections, and $%
[0,1]$ type connections are thus all trivial.

\subsection{The complex manifold specified by the general linear group $GL(n,%
\not C)$}

It seems that the expression in section IV is only developed for real
geometry, but in fact it also pertains to complex geometry. Here we iterate
the results in terms of complex geometry.

First of all, let's assume that we have chosen a metric for the complex
space,
\begin{equation}
A({ \bar \psi },{ \psi })=A_{\bar \alpha \beta }d{ \bar \psi }%
^\alpha d{ \psi }^\beta \text{ .}  \tag{5.5}
\end{equation}
It is an additional requirement to real geometry that the indices
$\alpha $ and $\beta $ belong to two different types, one is
normal component, the other is its complex conjugate, $\alpha
\rightarrow \bar \alpha $. The reason why we choose this
particular metric form will be explained in the next section.
Following the process of the last section, let's make the
connection preserve the metric $A( \bar \psi,\psi)$. Having chosen
the $\frac \partial {\partial { \bar \psi }^\alpha }$,$\frac
\partial {\partial {\frak \psi }^\beta }$ as basis $\{e_{\bar \alpha
},e_\beta \}$, the result is easily obtained from Eq.(4.19) that
\begin{equation}
dA_{\alpha \bar \beta }-A_{\gamma \bar \beta }\Gamma _{\;\alpha
}^\gamma -A_{\alpha \bar \gamma }\Gamma _{\;\bar \beta }^{\bar
\gamma }=0\text{ .} \tag{5.6}
\end{equation}
In obtaining the above equation, we have used the following equations due to
the parallel displacement
\begin{equation}
d\xi ^{\bar \alpha }+\Gamma _{\;\bar \gamma }^{\bar \alpha }\xi
^{\bar \gamma } =0\text{ ,}  \tag{5.7a}
\end{equation}
\begin{equation}
d\eta ^\beta +\Gamma _{\;\gamma }^\beta \eta ^\gamma =0\text{ ,}
\tag{5.7b}
\end{equation}
where $\xi ^{\bar \alpha }\ $and $\eta ^\beta $ are components for
the vector $X$ and $Y,$ $X=\xi ^{\bar \alpha }e_{\bar \alpha },$
$Y=\eta ^\beta e_\beta $. In analogy with the form (4.22),
Eq.(5.6) can be expressed in the following matrix form
\begin{equation}
dA=\Gamma ^t\cdot A+A\cdot \bar \Gamma  \tag{5.8}
\end{equation}
where $\bar \Gamma $ refers to the matrix with components $\Gamma _{\;\bar
\alpha }^{\bar \gamma }$. Performing exterior product on the above equation
turns the left hand trivial, and
\begin{eqnarray}
0 &=&ddA=d\Gamma ^t\cdot A-\Gamma ^t\wedge dA+dA\cdot \bar \Gamma +A\cdot d%
\bar \Gamma  \nonumber \\
\ &=&d\Gamma ^t\cdot A-\Gamma ^t\wedge (\Gamma ^t\cdot A)-\Gamma ^t\wedge
(A\cdot \bar \Gamma )  \nonumber \\
&&\ \ \ \ \ \ \ \ \ \ \ \ +(\Gamma ^t\cdot A+A\cdot \bar \Gamma )\wedge \bar
\Gamma +A\cdot d\bar \Gamma  \nonumber \\
\ &=&\Omega ^t\cdot A+A\cdot \bar \Omega ,      {(5.9)}\nonumber
\end{eqnarray}
where $\Omega ^t=d\Gamma ^t-\Gamma ^t\wedge \Gamma ^t$, in which the minus
sign comes from the transpose of matrices. Eq. (5.9) suggests the curvature
matrix and metric matrix are of anticommute. In Eq.(5.8), if $\Gamma ^t$ is
the $[1,0]$ type, then $\bar \Gamma $ is the $[0,1]$ type. Bearing in mind
that $d=\partial +\bar \partial $, Eq.(5.8) gives rise to
\begin{equation}
\partial A=\Gamma ^t\cdot A\text{ ,}  \tag{5.10a}
\end{equation}
i.e.
\begin{equation}
\Gamma =(A^t)^{-1}\cdot \partial A^t\text{ .}  \tag{5.10b}
\end{equation}
The inverse of matrix $A$ is so defined that
\begin{equation}
A_{\alpha \bar \beta }A^{\bar \beta \gamma }=\delta _\alpha ^{\;\,\gamma
},\;A^{\bar \alpha \gamma }A_{\gamma \bar \beta }=\delta _{\;\;\bar \beta }^{%
\bar \alpha }\text{ .}  \tag{5.11}
\end{equation}
Then due to (5.10b) the component form of $\Gamma $ is
\begin{equation}
\Gamma _{\;\alpha }^\beta =A^{\beta \bar \gamma }\frac{\partial
A_{\bar \gamma \alpha }}{\partial { z}^i}d{ z}^i\text{ .}
\tag{5.12}
\end{equation}
Consequently the curvature form (4.23) and (4.24) certainly hold, with the
variables now being complex
\begin{equation}
\Omega =(\Omega _\alpha ^\beta )=d\Gamma -\Gamma \wedge \Gamma
\text{ ,} \tag{4.23}
\end{equation}
from $\Gamma _{\;\alpha }^\beta =\Gamma _{k\alpha }^\beta d{
z}^k$, and
taking into account the definition of the components form of curvature, $%
\Omega _\alpha ^\beta =R_{\alpha jk}^\beta d{ \bar z}^j\wedge d{ z}^k $%
, it is straightforward to obtain
\begin{equation}
R_{\alpha jk}^\beta =\Gamma _{k\alpha ,j}^\beta -\Gamma _{j\alpha
,k}^\beta +\Gamma _{j\gamma }^\beta \Gamma _{k\alpha }^\gamma
-\Gamma _{k\gamma }^\beta \Gamma _{j\alpha }^\gamma \text{ .}
\tag{4.24}
\end{equation}
Here the meaning of $d{ \bar z}^j$ is the same as that of $d{ \bar
\psi }^j$, and likewise for $d{ z}^k$. From this form it is
obvious that the curvature is antisymmetric with respect to the
last two indices $j$, $k$. Except the requirement Eq. (5.4) due to
analytic condition, the manifold in virtue of $GL(n,\not C)$ adds
no more constraint to the connection.

\subsection{The complex manifold specified by the unitary group $U(n,\not C)$%
}

On the basis of the above general complex manifold, if we demand the metric
matrix $A$ satisfy
\begin{equation}
(A^t)^{*}=A  \tag{5.13}
\end{equation}
with components form as $A_{\bar \alpha \beta }=\bar A_{\alpha \bar \beta }$
i.e. if $A$-matrix is Hermitian conjugate to itself, then the metric is
called the Hermitian metric. All the components of the Hermitian metric form
the Hermitian matrix (Hermitian operator). This kind of operator is well
known in quantum mechanics and has many good properties. To keep the
Hermitian metric $d{ \bar \psi }^\alpha A_{\bar \alpha \beta }d{ \psi }%
^\beta $ and corresponding curvature invariant, the transformation group
must be a Unitary group $U(n,\not C)$.

\subsection{The K\"ahler manifold}

Additionally, as for an Hermitian manifold, if the connection $\Gamma
_{\alpha \gamma }^\beta $ is symmetric with respect to the indices $\alpha $
and $\gamma $, the manifold then is called K\"ahler manifold. The
transformation group for K\"ahler manifold is $SU(n,\not C)$.

\subsection{Independent components of curvature and Ricci tensor}

To summarize the above results relevant to curvature, we note that it
subjects to three constraints as the following.

(I) As shown in Eq. (5.4), the form of connection like $\Gamma _{\alpha \bar
\beta }^\gamma $ and $\Gamma _{\alpha \beta }^{\bar \gamma }$ is trivial,

(II) We only use [1, 0] type connection, so the [0, 1] type connection $%
\Gamma _{\bar \alpha \beta }^\gamma =\Gamma _{\bar \alpha \bar \beta }^{\bar
\gamma }=0$,

(III) The curvature form is [1, 1] type, so the forms $R_{\delta \alpha
\beta }^\gamma $ or $R_{\delta \bar \alpha \bar \beta }^\gamma $ vanish.

Under these constraints, one can calculate the following curvature
components one by one according to the explicit form (4.24): $R_{\bar \beta
jk}^\alpha $, $R_{\beta \bar jk}^\alpha $, $R_{\beta j\bar k}^\alpha $, $R_{%
\bar \beta \bar jk}^\alpha $, $R_{\bar \beta j\bar k}^\alpha $, $R_{\beta
\bar j\bar k}^\alpha $, $R_{\bar \beta \bar j\bar k}^\alpha $, $R_{\bar \beta
jk}^{\bar \alpha }$, $R_{\beta \bar jk}^{\bar \alpha }$, $R_{\beta j\bar k}^{%
\bar \alpha }$, $R_{\bar \beta \bar jk}^{\bar \alpha }$, $R_{\bar \beta j%
\bar k}^{\bar \alpha }$. Finally one concludes that only four of them are
nonzero
\begin{equation}
R_{\beta \bar jk}^\alpha =\Gamma _{k\beta ,\bar j}^\alpha \text{
,} \tag{5.14a}
\end{equation}
\begin{equation}
R_{\beta j\bar k}^\alpha =-\Gamma _{j\beta ,\bar k}^\alpha \text{
,} \tag{5.14b}
\end{equation}
\begin{equation}
R_{\bar \beta \bar jk}^{\bar \alpha } =\Gamma _{k\bar \beta ,\bar
j}^{\bar \alpha }\text{ ,}  \tag{5.14c}
\end{equation}
\begin{equation}
R_{\bar \beta j\bar k}^{\bar \alpha } =-\Gamma _{j\bar \beta ,\bar k}^{%
\bar \alpha }\text{ .}  \tag{5.14d}
\end{equation}
The above four equations hold independent of the torsion [The
torsion is defined as $T_{\gamma \beta }^\alpha =\Gamma _{\gamma
\beta }^\alpha -\Gamma _{\beta \gamma }^\alpha $] since the
evaluating process has nothing to do with it. When the torsion is
absent, the first two curvatures in Eq. (5.14)
display more symmetries $R_{\beta \bar jk}^\alpha =R_{k\bar j\beta }^\alpha $%
, $R_{\beta j\bar k}^\alpha =R_{j\beta \bar k}^\alpha $.

The constraints certainly affect the Ricci tensor. Based on Eq.(5.14), the
Ricci tensors are obtained by tracing $\Omega $ in Eq. (4.23),

\begin{eqnarray}
&&\ R_{\beta \bar jk}^\alpha \stackrel{\alpha =\beta }{\longrightarrow }R_{%
\bar jk}\text{,}\;R_{\beta j\bar k}^\alpha \stackrel{\alpha =\beta }{%
\longrightarrow }R_{j\bar k}\text{,}  \nonumber \\
&&\ R_{\bar \beta \bar jk}^{\bar \alpha }\stackrel{\bar \alpha =\bar \beta }{%
\longrightarrow }R_{\bar jk}\text{,}\;R_{\bar \beta j\bar k}^{\bar \alpha }%
\stackrel{\bar \alpha =\bar \beta }{\longrightarrow }R_{j\bar
k}\text{.}      {5.15}  \nonumber
\end{eqnarray}
The equation suggests only two independent forms of Ricci tensor, $R_{\bar j%
k}$ and $R_{j\bar k}$, exist. Furthermore, the curvature component $R_{\beta
\bar jk}^\alpha $ is antisymmetric with respect to the indices $\bar j$ and $%
k$, i.e. $R_{\bar jk}=-R_{k\bar j}$, whence only one form of Ricci tensor
would occur. We will return to this antisymmetry in section IX.

Now the Ricci tensor defined by (5.15) can be explicitly written as

\begin{equation}
R_{\bar \alpha \beta }=R_{\gamma \bar \alpha \beta }^\gamma =\Gamma _{\beta
\gamma ,\bar \alpha }^\gamma (\text{or }\Gamma _{\beta \bar \gamma ,\bar
\alpha }^{\bar \gamma })=\frac{\partial \Gamma _{\beta \gamma }^\gamma }{%
\partial { \bar z}^\alpha }=\frac{\partial ^2\ln A}{\partial { \bar z}%
^\alpha { z}^\beta }\text{ ,}  \tag{5.16}
\end{equation}
here the notation $A$ stands for determinant $\det (A_{\bar \alpha \beta })$%
, and the last step is obtained by applying the differential relation of a
matrix determinant
\begin{equation}
A_{,\nu }=A\text{ }A^{\beta \bar \alpha }A_{\bar \alpha \beta ,\nu
} \tag{5.17}
\end{equation}
to the derivatives of $\Gamma _{\beta \gamma }^\gamma $%
\begin{equation}
\frac{\partial \Gamma _{\beta \gamma }^\gamma }{\partial { \bar
z}^\alpha
}=\frac \partial {\partial { \bar z}^\alpha }(A^{\gamma \bar \rho }\frac{%
\partial A_{\bar \rho \gamma }}{\partial { z}^\beta })=\frac \partial {%
\partial { \bar z}^\alpha }(\frac \partial {\partial { z}^\beta }\ln A)%
\text{ .}  \tag{5.18}
\end{equation}

The above discussions set no limit on transformation group, hence pertain to
general situations, say, $GL(n,\not C)$. For the Hermitian or K\"ahler
manifold, the independent components of curvature and Ricci tensor should
decrease. If $A_{\bar \alpha \beta }$ is Hermitian, then $A_{\bar \alpha
\beta }^{*}=A_{\beta \bar \alpha }$ and $\det (A_{\bar \alpha \beta })=\det
(A_{\bar \alpha \beta }^{\dagger })$, applying (5.14a), (5.14d) or (5.14b),
(5.14c) to (5.18), the following expression can be achieved
\begin{equation}
R_{\bar \alpha \beta }^{*}=R_{\alpha \bar \beta }\text{ .}
\tag{5.19}
\end{equation}
The specification of K\"ahler manifold adds no more constraint to the form
of Ricci tensor except the symmetric indices of connection in equation
(5.14a), (5.14b).

Now let's count the number of the independent components of the Ricci
tensor. Apart from the fact that the two indices are antisymmetric, the
group $GL(4,\not C)$ adds no constraints. So in this case the set $\{\bar
\alpha ,\beta \}$, with $\alpha ,\beta $ being over $1,2,3,4$, has totally $%
16$ elements. While $\{\alpha ,\bar \beta \}$ need not be taken into account
for antisymmetry. The freedom of $A_{\bar \alpha \beta }$, however, is
totally $32$ components, so it requires $16$ gauge conditions to solve the
field Eq. (9.8). Furthermore, if the manifold is Hermitian, i.e. the
Eq.(5.19) holds, $R_{\bar \alpha \beta }^{*}=-R_{\bar \beta \alpha }$, then
the elements of set $\{\bar \alpha ,\beta \}$ decrease to $6$, and the
freedom of $A_{\bar \alpha \beta }$ also shrinks to $20$, hence $14$
additional equations are needed to resolve the $A_{\bar \alpha \beta }$,
almost the same as that of $GL(4,\not C)$.

\section{The Metric for Fermion Field}

A quantum fermion field is customarily expressed by Dirac spinor
${ \psi (x)}$, and the inner product of ${ \psi (x)}$ is
prescribed as
\begin{equation}
{ \bar \psi }{ (}x{ ){ \psi }(}x)={ \psi }^{\dagger }%
{ (}x{ )\gamma }_0{ \psi }{ (}x{ )}\text{ ,} \tag{6.1}
\end{equation}
where
\begin{equation}
{ \gamma }_0=\left(
\begin{array}{cccc}
1 & 0 & 0 & 0 \\
0 & 1 & 0 & 0 \\
0 & 0 & -1 & 0 \\
0 & 0 & 0 & -1
\end{array}
\right) \text{ .}  \tag{6.2}
\end{equation}
To be consistent with conventional quantum field theory (CFQT), here the $%
{ \bar \psi }{ (}x{ )}$ is defined as ${ \psi }^{\dagger }%
{ (}x{ )\gamma }_0$, which is different from previous denotation $%
{ \bar \psi }_\alpha { (}x{ )}={ \psi }_\alpha ^{*}{ (}x%
{ )}$. Henceforth we distinguish the difference by whether or not
the subscripts or superscripts are used: if used, then the latter
definition is
available, otherwise the former definition holds. The former definition $%
{ \bar \psi }{ (}x{ )=}{ \psi }^{\dagger }{ (}x{ %
)\gamma }_0$ is meaningful in making the inner product ${ \bar \psi }%
{ (}x{ ){ \psi }(}x)$ invariant under the transformation group $%
SL(2,\not C)$, which is the spinor-representation of Lorentz group. The
invariance is a direct corollary of Lorentz invariance of Dirac equation [2].

To apply the differential geometry, let's generalize the inner product. To
do so the definition of metric should be valid only within a very small
region of the complex space, e.g. the inner product ${ \bar \psi }{ (%
}x{ ){ \psi }(}x)$ should be generalized to $d{ \bar \psi }{ %
(}x{ )}d{ \psi }{ (}x)$, and only in this sense does the inner
product remain invariant under the transformation of group
$SL(2,\not C)$.
Any infinitesimal transformation is now performed on $d{ \psi }{ (}%
x) $ instead of on ${ \psi }{ (}x)$. $d{ \psi }{ (}x)$ can be
viewed as plane wave locally and ${ \psi }{ (}x)$ has conventional
meaning on a larger scale.

The above inner product holds when there is no interaction, if any
interaction arises, the product has to be interpreted by a general
metric form $A_{\alpha \bar \beta }d{ \psi }^\alpha d{ \psi
}^{*\beta }$. The above form $d{ \bar \psi }{ (}x{ )}d{ \psi }{
(}x)$ is
only a special case while $A_{\alpha \beta }=\eta \,\delta _{\alpha \beta }$%
, $\eta =1$ for $\alpha =1,2$ and $\eta =-1$ for $\alpha =3,4$. In
conclusion, the metric of the quantum field is demanded to be
\begin{equation}
A({ \bar \psi },{ \psi })=A_{\alpha \bar \beta }d{ \psi }^\alpha d%
{ \psi }^{*\beta }=d{ \psi }^\alpha A_{\alpha \bar \beta }d{ \bar
\psi }^\beta \text{ ,}  \tag{6.3}
\end{equation}
which does not violate experiences from QED.

\section{Two types of curving in quantum fields}

In a two-dimension (2-D) plane, the curves of hyperbolic or elliptic type
can be interpreted by the equations like $a^2\,x^2-b^2\,y^2=1$ or $%
a^2\,x^2+b^2\,y^2=1$. General linear transformations in 2-D [Known to be $%
GL(2,R)$ group. They have the general form $\left(
\begin{array}{cc}
c_{11} & c_{12} \\
c_{21} & c_{22}
\end{array}
\right) $, with $c_{ij}$ being real numbers] can't change the types of
curves, since the performance of linear transformations has two identical
manners: one is to change the objective curves; the other is to change the
coordinate axes. Applying the latter manner, the types of the curves are
obviously reserved. In other words, under general linear transformations
within 2-D, if a curve in 2-D is ever a type, it will forever be the type.
In this paper we take into account only the quadratically homogeneous forms
for variables, so considering these two types is enough.

The above argument is expected to hold also in four dimension space (4-D).%
{\bf \ }It becomes complicate in this case, since a 4-D curve being
hyperbolic in one projected plane may be an ellipse in another. So the
definition of hyperbolic or elliptic type has to be extended: For a 4-D
curve, if there exists at least one 2-D plane so that the curve projected
into it is hyperbolic, then curve is called hyperbolic, otherwise it is
elliptic. In our concerns the 4-D interval [In this paper ''metric'' means
the same as ''interval''] $c^2t^2-x^2-y^2-z^2$ in special relativity is
hyperbolic type, because if the equation $c^2t^2-x^2-y^2-z^2=constant$ is
projected into any 2-D subspace including the time axis, the resultant curve
is a hyperbolic type. Lorentz group preserves the interval $%
c^2t^2-x^2-y^2-z^2$, so the hyperbolic characteristic is also the main
feature of Lorentz group, which is called noncompact in terminology of group
theory. In contrast, the group $SO(4)$ is compact for it preserve the
interval $c^2t^2+x^2+y^2+z^2$, which is a elliptic type. Instead of
requiring the transformation groups to preserve the interval, now we only
require the types of interval to be preserved. Then we can find that
hyperbolic characteristic of the interval $c^2t^2-x^2-y^2-z^2$ will not
change under the general linear group $GL(4,R)$, similar to the above 2-D
case. Generally, to keep the type of an $n$-D interval of quadratic
homogeneous form, that the transformation is linear is sufficient. Here and
hereafter the explanations of concepts relevant to Group are rough and only
for later application.

The above definitions of hyperbolic and elliptic type in 2-D are for the
global space. In a local region of the whole space, the two types should be
interpreted by $a^2\,dx^2-b^2\,dy^2=1$ and $a^2\,dx^2+b^2\,dy^2=1$
respectively. Correspondingly, if Special Relativity is treated locally,
then the metric is
\begin{equation}
c^2dt^2-dx^2-dy^2-dz^2\text{ .}  \tag{7.1}
\end{equation}
In a small local region, the infinitesimal transformations of $GL(4,R)$
cannot change the types of the curves. Therefore, even though a
transformation changes (7.1) to a general form $g^{\alpha \beta }dx_\alpha
dx_\beta $, by which we cannot tell its type directly, the type of the
metric $g^{\alpha \beta }dx_\alpha dx_\beta $ remains the same as Eq. (7.1).

Now we extend the above discussions and knowledge to complex space. In
following three paragraphs, we construct the {\bf parallelisms} between the
metrics of real space and complex space, as well as the {\bf parallelisms}
between the groups preserving them, then generalize our understanding of
hyperbolic and elliptic type in real space to the understanding of complex
space.

Let's consider the complex-space metrics defined in the last section, ${ %
\psi }^{\dagger }{ (}x{ )\gamma }_0{ \psi }{ (}%
x)=u_1^{*}u_1+u_2^{*}u_2-u_3^{*}u_3-u_4^{*}u_4$, where ${ \psi }{ (}x%
{ )}$ is a complex spinor ${ \psi }^{\dagger
}=(u_1^{*},u_2^{*},u_3^{*},u_4^{*})$. This metric is invariant
under the transformations of group $SL(2,\not C)$, $\not C$ means
that the elemental variables in the group are complex ones. [3,
4]. It can be verified [3] that
there exists a two-to-one mapping between all the elements of the group $%
SL(2,\not C)$ and all the elements of the proper, orthochronous Lorentz
group. Hereby the group $SL(2,\not C)$ is viewed as {\bf complex parallellism%
} of the Lorentz group (which is born from physics in real space-time).

Now let's elucidate the relationship between group $SL(2,\not C)$ and
aforementioned metric ${ \psi }^{\dagger }{ (}x{ )\gamma }_0%
{ \psi }{ (}x)$. The quantities which are transformed according to
lowest-dimensional nontrivial representation of $SL(2,\not C)$ are
called two-component {\it spinors}, which are doublets as states
for spin. On the basis of higher dimensional representation of
$SL(2,\not C)$ other spinors with more components can be
constructed. In our case the four-component
spinor ${ \psi }{ (}x{ )}$ is transformed according to named $%
D^{(\frac 12,\frac 12)}$ representation of $SL(2,\not C)$ [involving $\gamma
$-matrices], their relationship can be derived from the covariance of Dirac
equation and we omit all the relevant details here. Also it can be confirmed
that the metric form ${ \psi }^{\dagger }{ (}x{ )\gamma }_0{ %
\psi }{ (}x)$ remains unaltered under the transformation of $D^{(\frac 1%
2,\frac 12)}$ representation of $SL(2,\not C)$. In a word, the metric ${ %
\psi }^{\dagger }{ (}x{ )\gamma }_0{ \psi }{ (}x)$ is the {\bf
complex parallellism} to the real interval $c^2t^2-x^2-y^2-z^2$.
Consequently the {\bf complex parallelism} to Eq. (7.1) should be
\begin{equation}
d{ \bar \psi }{ (}x{ )}d{ \psi }{ (}x)\text{ .} \tag{7.2}
\end{equation}
Now we know ${ \psi }^{\dagger }{ (}x{ )\gamma }_0{ \psi }%
{ (}x)$--and hence $d{ \bar \psi }{ (}x{ )}d{ \psi }%
{ (}x)$--is hyperbolic. However, we don't know what is the largest
group--which includes $SL(2,\not C)$ as a subgroup--that makes the
type of
metric ${ \psi }^{\dagger }{ (}x{ )\gamma }_0{ \psi }{ (%
}x)$ unchanged. Searching for what is exactly the largest group would be a
tedious work and might simultaneously deviate us from the main line of
developing this theory. In this paper we simply assume the latest group is $%
GL(4,\not C)$ (in fact must be a subgroup of $GL(4,\not C)$) and bearing in
mind that it is for hyperbolic type.

The {\bf metric parallelism} between elliptic types of real space and
complex space can be constructed by directly generalizing each of real axes
to a complex one. We are familiar with the metric defined in the 3-D space $%
x^2+y^2+z^2${\bf ,} which is invariant under the rotation group $SO(3)$.
Suppose a complex space spanned by wave functions with only three
components, $({ \psi }_1{ (}x),{ \psi }_2{ (}x),{ \psi }_3%
{ (}x))$, corresponding to metric $x^2+y^2+z^2${\bf \ }and
consistent with the probability in quantum mechanics, we conclude
that its metric
should be of the form ${ \psi }_1^{*}{ (}x){ \psi }_1{ (}x)+%
{ \psi }_2^{*}{ (}x){ \psi }_2{ (}x)+{ \psi }_3^{*}{ %
(}x){ \psi }_3{ (}x)$, where the integration over the
configuration space is implied. To preserve the probability in the
whole complex space, the quantum mechanics requires any
transformations performed on the wave function $({ \psi }_1,{ \psi
}_2,{ \psi }_3)$ should be elements of special unitary group
$SU(3)$ or its subgroups. $SU(n)$ groups are compact,
just as $SO(n)$. Hereby the metric ${ \psi }_1^{*}{ (}x){ \psi }_1%
{ (}x)+{ \psi }_2^{*}{ (}x){ \psi }_2{ (}x)+{ \psi }%
_3^{*}{ (}x){ \psi }_3{ (}x)$ and group $SU(n)$ are viewed as the
{\bf complex parallellism} to that of elliptic type of real space.
The caution should be practiced that it is not perfect to
generalize the rotation transformation in real space directly to
conservation of angular momentum as done in quantum mechanics to
achieve the same result that metric
${ \psi }_1^{*}{ (}x){ \psi }_1{ (}x)+{ \psi }_2^{*}%
{ (}x){ \psi }_2{ (}x)+{ \psi }_3^{*}{ (}x){ \psi }_3%
{ (}x)$ and $SU(n)$ are elliptic type. Since the Lorentz group
includes a spatial rotation part besides the boost part, from the
Noether theorem of quantum field theory one can also derive the
conservation of angular momentum. Then there will be confusions
between the two types. The degrees of freedom of elliptic type
presented in this paper are those degrees of freedom such as
electro-charges, isospins, or colours etc., other than spins and
spatial angular momentum. The largest group to preserve the
elliptic
type of the metric $\sum\nolimits_i{ \psi }_i^{*}{ (}x){ \psi }_i%
{ (}x)$ is certainly $U(n)$ groups.

We have endowed the Lorentz symmetry and special Unitary symmetry with
hyperbolic type and elliptic type, the corresponding types of metric defined
according to these two symmetries remain stable even after the groups are
appropriately extended. We know $U(n)$ group will not change the type of
metric $\sum\nolimits_i{ \psi }_i^{*}{ (}x){ \psi }_i{ (}x)$%
, and a subgroup of $GL(4,\not C)$ will not change the type of ${ \psi }%
^{\dagger }{ (}x{ )\gamma }_0{ \psi }{ (}x)$. It is the extended
groups that provide more degrees of freedom over which the complex
space spanned by wave functions is curved. Corresponding to the
two symmetries, there are two types of curvings. Since for a
fermion, such as quark, it has both the curving
characteristics---hyperbolic type and elliptic type---spinor and
colour, we should combine the two types of metric into one general
form
\begin{equation}
A_{\alpha \beta }^{ac}d{ \bar \psi }_a^\alpha d{ \psi }_c^\beta
\text{.}  \tag{7.3}
\end{equation}
where we generalize ${ \psi }^\alpha { (}x)$ to ${ \psi }%
_c^\alpha { (}x),c=1,2,3$ are colour indices, $\alpha $, $\beta
$-are spinor indices. The metric matrix $(A_{\alpha \beta }^{ac})$
may be written separately as a hyperbolic part multiplied by an
elliptic part.

\section{Motion equation for fermions}

\subsection{General motion equation for fermions}

After reviewing the geometrical method, let's turn to its application to
physical problems.

As an equal level observer, assumedly an electron can test another electron
without losing any information. To an observer what is the motion equation
of the observed electron? For an observer in space-time the right equation
is certainly the Schr\"odinger equation or the Dirac equation. But as an
equal level observer, it constructs the equation as follows: there always
exists a complex local frame for the observer in which the observed electron
looks like a plane wave (a free electron), in spite of the existence of
interaction--similar to that happens in General Relativity--saying that
there always exists a local frame in space-time in which the observed
particle looks like a free one moving along a straight line, in spite of the
gravitation. In terms of geometry, it means that the motion of electron is
just a geodetic line in complex space, i.e., the parallel transport
(displacement).

Conventionally a free electron is described by plane wave $u(\vec p%
)\,e^{ip\cdot x}$, $u(\vec p)$ is Dirac spinor. The complex space here
should be of 4-dimension since $u(\vec p)$ possesses four components,
furthermore, it should be hyperbolic as argued in the preceding section. In
this respect the transformation group for the space should be $GL(4,\not C)$%
. Since the complex space is continuously spanned by components of
local spinor $\{{ \psi }^1,{ \psi }^2,{ \psi }^3,{ \psi }^4\}$,
its
natural bases are $\{e_\alpha =\frac \partial {\partial { \psi }^\alpha }%
,e_{\bar \alpha }=\frac \partial {\partial { \bar \psi }^\alpha
}\}$.
Consequently, $X=\xi ^{\bar \alpha }e_{\bar \alpha }=d{ \bar \psi }%
^\alpha \frac \partial {\partial { \bar \psi }^\alpha }$, then the
parallel displacement (5.6a) $(d\xi ^{\bar \alpha }+\Gamma
_{\;\bar \gamma }^{\bar \alpha }\xi ^{\bar \gamma })=0$ turns out
to be
\begin{equation}
dd{ \bar \psi }^\alpha +\Gamma _{\;\bar \gamma }^{\bar \alpha }d{
\bar \psi }^\gamma =0\text{.}  \tag{8.1}
\end{equation}
Similarly, another equation for ${ \psi }^\alpha $ holds,
\begin{equation}
(dd{ \psi }^\beta +\Gamma _{\;\gamma }^\beta d{ \psi }^\gamma )=0\text{%
.}  \tag{8.2}
\end{equation}
Above two equations are the general forms of motion equation for fermions.

\subsection{ The rule of calculating the differential $d$}

In Eq. (8.2) the component $d{ \psi }^\beta $ is an infinitesimal
change of ${ \psi }^\beta $ relative to its neighboring points,
which is determined by ${ \psi }^\beta $ and the transformation
performed on it. Suppose that after a transformation the wave
function ${ \psi }^\beta $ changes to ${ \psi }^{\prime \beta },$
then resembling the form of Eq.(3.7), the wave function ${ \psi
}^{\prime \beta }$ can be expressed by variables ${ \psi }^\beta
$,
\begin{equation}
{ \psi }^{\prime \beta }={ \psi }^{\prime \beta }({ \psi }^1,{
\bar \psi }^1;{ \psi }^2,{ \bar \psi }^2;\cdots ,{ \psi }^n,{ \bar
\psi }^n)\text{ .}  \tag{8.3}
\end{equation}
Now the differential form $d{ \psi }^\beta $ can be written
explicitly as
\begin{equation}
d{ \psi }^{\prime \beta }=\frac{\partial { \psi }^{\prime \beta }}{%
\partial { \psi }^\gamma }d{ \psi }^\gamma +\frac{\partial { \psi }%
^{\prime \beta }}{\partial { \bar \psi }^\gamma }d{ \bar \psi
}^\gamma \text{ ,}  \tag{8.4}
\end{equation}
The derivative $\frac{\partial { \psi }^{\prime \beta }}{\partial { %
\psi }^\gamma }$ in the above equation however, is not operable in practical
calculation.

Practically, we want to know how far the Eqs. (8.1), (8.2) deviate
from the conventional Dirac equation. Or to put it alternatively,
how can the terms in Dirac equation be finally restored in local
region by reducing aforementioned parallel displacement? To obtain
a form of Dirac equation, let's first recall the replacement used
in GR: $d^2{ x}^\mu +\Gamma
_{\,\;\sigma }^\mu d{ x}^\sigma =0\rightarrow \frac{d^2{ x}^\mu }{dt^2}%
+\Gamma _{\rho \sigma }^\mu \frac{d{ x}^\rho }{dt}\frac{d{ x}^\sigma }{%
dt}=0$. That reminds us to replace the differential operator $d$ in Eqs.
(8.1), (8.2) by some forms of derivatives with respect to space-time. In
view of the quadratic form of Eqs. (8.1), (8.2), it is helpful to know the
following quadratic form [2] of Dirac equation before we do some replacement
of operator $d$,
\begin{equation}
\lbrack (i\partial _\mu -eA_\mu )(i\partial ^\mu -eA^\mu )-\frac
i2e\gamma _\mu \gamma _\nu F^{\mu \nu }]{ \psi }=m^2{ \psi }\text{
,} \tag{8.5}
\end{equation}
where $F^{\mu \nu }=\partial _\mu A_\nu -\partial _\nu A_\mu $ is the field
tensor of common sense. In what follows, we put $e=1$. Then the explicit
form of Eq. (8.5) is
\begin{equation}
\partial _\mu \partial ^\mu { \psi +}i\,\partial _\mu A^\mu { \psi
+}i\,A_\mu \partial ^\mu { \psi -}\,A_\mu A^\mu { \psi }+\frac i2%
\,\gamma _\mu \gamma _\nu F^{\mu \nu }{ \psi }=-m^2{ \psi }\text{ ,%
}  \tag{8.6a}
\end{equation}
By using a weaker Lorentz condition $\partial _\mu A^\mu { \psi
}=0$ of Gupta and Bleuler [5] instead of the original one
$\partial _\mu A^\mu =0$, the above equation changes to
\begin{equation}
\Box { \psi +}i\,\,A_\mu \partial ^\mu { \psi -}\,A_\mu A^\mu
{ \psi }+\frac i2\,\gamma _\mu \gamma _\nu F^{\mu \nu }{ \psi }%
=-m^2{ \psi }\text{ .}  \tag{8.6b}
\end{equation}
As a reasonable approximation to leading-order QED, the quadratic terms of $%
A^\mu $, such as $A_\mu A^\mu { \psi }$ in the above equation and
similar terms in our following calculation will be temporarily
omitted.

Now let's turn to treating the equation (8.2). First let's
substitute the
explicit form of $\Gamma _{\beta \gamma }^\alpha $, $A^{\alpha \bar \rho }%
\frac{\partial A_{\bar \rho \gamma }}{\partial { z}^\beta }$, into
the equation (8.2), concerning only the second term, it yields
\begin{eqnarray}
\Gamma _{\beta \gamma }^\alpha d{ \psi }^\beta d{ \psi }^\gamma
&=&A^{\alpha \bar \rho }\frac{\partial A_{\bar \rho \gamma }}{\partial { %
\psi }^\beta }d{ \psi }^\beta d{ \psi }^\gamma  \nonumber \\
\ &=&A^{\alpha \bar \rho }dA_{\bar \rho \gamma }d{ \psi }^\gamma
\text{ .}        {((8.7)}  \nonumber
\end{eqnarray}
Then the form of motion equation (8.2) becomes
\begin{equation}
dd{ \psi }^\alpha +A^{\alpha \bar \rho }\,dA_{\bar \rho \gamma }\,d{ %
\psi }^\gamma =0\text{.}  \tag{8.8}
\end{equation}
Extending the discussion of Eqs. (6.1)\symbol{126}(6.3) straightforwardly by
considering the vertex $\gamma ^\mu A_\mu $ used in QED, we can interpret
the metric tensor $A_{\bar \rho \gamma }$ with interaction as $(A_{\bar \rho
\gamma })_{4\times 4}=\gamma _0+\gamma _0\gamma ^\mu A_\mu $ (Henceforth we
continue to use $A^\mu $ to express the field potential. Take care not to
confuse it with the metric tensor $A^{\alpha \beta }$). It is easy to verify
that $(A^{\alpha \bar \rho })_{4\times 4}=\gamma _0-\gamma ^\mu \gamma
_0A_\mu $. To obtain a form of Dirac equation, comparing the second term of
Eq. (8.8) with the term $\gamma _\mu \gamma _\nu F^{\mu \nu }$ in Eq. (8.6),
we find it is effective to replace $d$ by $\gamma _\mu \partial ^\mu $.
Obviously, the replacement $d\rightarrow $ $\gamma _\mu \partial ^\mu $
performed to the first term of Eq. (8.8) directly induces $dd\rightarrow
\Box $.

Furthermore, by analyzing the dimensions of Eq. (8.8), we find the
dimensions of the first term and that of the second term are not
equal. In the natural units, the first term of the above equation
has the dimension of energy square if assuming ${ \psi }$
dimensionless. As for the second term, the form $A^{\alpha \bar
\rho }=\gamma _0-\gamma ^\mu \gamma _0A_\mu $ can be simplified as
$1+B$, and $dA_{\bar \rho \gamma }$ [$\gamma _\mu
\partial ^\mu (\gamma _\nu A^\nu )$] is energy square too, but the derivative
performed on ${ \psi }$ also contribute the dimension of an
energy, so an extra energy dimension exists. To remedy the{\bf \
}unequal dimensions of two sides and according to our experience
of treating the Schr\"odinger equation with a nonlocal
interaction-potential [6], we add a line integral with respect to
space-time $x^\mu =(t,\,\vec x)$ to the second term. The
integration is accompanied by the purely imaginary number
$i\,$according to requirement restoring the Dirac equation. Now
the form of equation (8.8) becomes
\begin{equation}
dd{ \psi }^\alpha +i\,\int_{x^\mu (-\infty )}^{x^\mu
(t)}\,A^{\alpha \bar \rho }\,dA_{\bar \rho \gamma }\,d{ \psi
}^\gamma dx=0\text{ ,} \tag{8.9}
\end{equation}
Consistently, the infinitesimal integral measurement $dx$ should also be
linked with $\gamma $-Matrix, assumed as $dx\rightarrow \gamma _\lambda
dx^\lambda $.

\subsection{Restoring the terms in Dirac equation}

Now let's evaluate the results of the second term $\int_{x^\mu (-\infty
)}^{x^\mu (t)}\,A^{\alpha \bar \rho }\,dA_{\bar \rho \gamma }\,d{ \psi }%
^\gamma dx$ in Eq. (8.9). It is convenient to write out the integral in
terms of matrix
\begin{eqnarray}
&&\ \ \int (1-\gamma _\mu A^\mu )\gamma _\sigma \partial ^\sigma (1+\gamma
_\nu A^\nu )\gamma _\rho \gamma _\lambda \partial ^\rho { \psi \,}%
dx^\lambda \text{ }  \nonumber \\
\ &=&\int \gamma _\sigma \partial ^\sigma (\gamma _\nu A^\nu )\gamma _\rho
\gamma _\lambda \partial ^\rho { \psi \,}dx^\lambda  \nonumber \\
&&\ \ -\int \gamma _\mu A^\mu \gamma _\sigma \partial ^\sigma
(\gamma _\nu A^\nu )\gamma _\rho \gamma _\lambda \partial ^\rho {
\psi \,}dx^\lambda \text{,}     {(8.10)}  \nonumber
\end{eqnarray}
here ${ \tilde \psi =}\{{ \psi }^1,{ \psi }^2,{ \psi }^3,{ %
\psi }^4\}$ is a four-component vector, $A^{\alpha \bar \rho }\,$ and $A_{%
\bar \rho \gamma }$ all in their matrix forms, $1-\gamma _\mu A^\mu $ and $%
1+\gamma _\nu A^\nu $. We will mainly deal with the first integral in
(8.10), since the second term includes two $A^\mu $ factors we temporarily
omit it as second-order perturbation. The calculation involves the formulae
of $\gamma -$matrices in simplifying the product $\gamma _\sigma \gamma _\nu
\gamma _\rho \gamma _\lambda $. The detailed evaluation on the first
integral is put to Appendix A, here we only show the result in (A. 7)
\begin{eqnarray}
&&\ \int \gamma _\sigma \partial ^\sigma (\gamma _\nu A^\nu )\gamma _\rho
\gamma _\lambda \partial ^\rho { \psi \,}dx^\lambda  \nonumber \\
\ &=&A_\nu \partial ^\nu { \psi +}\frac 12\gamma _\mu \gamma _\nu \tilde F%
^{\mu \nu }{ \psi +}\frac 12\gamma _\mu \gamma _\nu F^{\mu \nu }{ %
\psi }  \nonumber \\
&&\ +\frac 12\gamma _\rho \gamma _\lambda \int (\partial _\nu
A^\nu )(dx^\lambda \partial ^\rho -dx^\rho \partial ^\lambda ){
\psi \,}
\nonumber \\
&&\ +i\epsilon _{\sigma \rho \lambda \mu }\gamma _5\gamma ^\mu
\int \partial ^\sigma (\gamma _\nu A^\nu )\partial ^\rho { \psi
\,}dx^\lambda -\int
A_\nu \partial _\lambda \partial ^\nu { \psi \,}dx^\lambda -\frac 12%
\gamma _\mu \gamma _\nu \int \tilde F^{\mu \nu }\partial _\lambda
{ \psi
\,}dx^\lambda  \nonumber \\
&&\ -\frac 12\gamma _\mu \gamma _\nu \int \partial _\lambda F^{\mu \nu }{ %
\psi \,}dx^\lambda -\gamma _\mu \gamma _\nu \int \partial ^\sigma A^\mu
\partial _\sigma { \psi \,}dx^\nu         {(8.11)}  \nonumber
\end{eqnarray}
where $\tilde F^{\mu \nu }=A^\mu \partial ^\nu -A^\nu \partial ^\mu $.
Submitting the integral result back into (8.9) yields

\begin{eqnarray}
&&\Box { \psi +}i\,\,A_\mu \partial ^\mu { \psi }+\frac i2\,\gamma
_\mu \gamma _\nu F^{\mu \nu }{ \psi }{ +}\frac{i\,}2\gamma _\mu
\gamma _\nu \tilde F^{\mu \nu }{ \psi }  \nonumber \\
&&+\frac{i\,}2\gamma _\rho \gamma _\lambda \int (\partial _\nu
A^\nu )(dx^\lambda \partial ^\rho -dx^\rho \partial ^\lambda ){
\psi \,}
\nonumber \\
&&-\,\,\epsilon _{\sigma \rho \lambda \mu }\gamma _5\gamma ^\mu \int
\partial ^\sigma (\gamma _\nu A^\nu )\partial ^\rho { \psi \,}dx^\lambda
-i\,\int A_\nu \partial _\lambda \partial ^\nu { \psi \,}dx^\lambda -%
\frac{i\,}2\gamma _\mu \gamma _\nu \int \tilde F^{\mu \nu }\partial _\lambda
{ \psi \,}dx^\lambda  \nonumber \\
&&-\frac{i\,}2\gamma _\mu \gamma _\nu \int \partial _\lambda F^{\mu \nu }%
{ \psi \,}dx^\lambda -i\,\,\gamma _\mu \gamma _\nu \int \partial
^\sigma
A^\mu \partial _\sigma { \psi \,}dx^\nu  \nonumber \\
&=&0\text{ ,}     {(8.12)}   \nonumber
\end{eqnarray}
it is found that the terms $i\,A_\nu \partial ^\nu { \psi }{ \
}$and $\frac{i\,}2\gamma _\mu \gamma _\nu F^{\mu \nu }{ \psi }$
are just
those required by (8.6b). However, quite a few other terms like $\frac{i\,}2%
\gamma _\mu \gamma _\nu \tilde F^{\mu \nu }{ \psi }$, $-i\,\int
A_\nu
\partial _\lambda \partial ^\nu { \psi \,}dx^\lambda $, etc. accompany
the required ones. Some of the redundant terms, such as
$\frac{i\,}2\gamma _\rho \gamma _\lambda \int (\partial _\nu A^\nu
)(dx^\lambda \partial ^\rho -dx^\rho \partial ^\lambda ){ \psi
\,}$,$\,\epsilon _{\sigma \rho \lambda \mu }\gamma _5\gamma ^\mu
\int \partial ^\sigma (\gamma _\nu A^\nu )\partial ^\rho { \psi
\,}dx^\lambda $, $i\,\gamma _\mu \gamma _\nu \int \partial ^\sigma
A^\mu \partial _\sigma { \psi \,}dx^\nu $, with integrand relevant
to local angular momentum $(dx^\lambda \partial ^\rho -dx^\rho
\partial ^\lambda )$, which is assumed second--order small in perturbation
theory according to our numerical experience, can be omitted
temporarily. The contributions of remainder terms like
$\frac{i\,}2\gamma _\mu \gamma _\nu \tilde F^{\mu \nu }{ \psi }$,
$i\,\int A_\nu \partial _\lambda
\partial ^\nu { \psi \,}dx^\lambda $, $\frac{i\,}2\gamma _\mu \gamma _\nu
\int \tilde F^{\mu \nu }\partial _\lambda { \psi \,}dx^\lambda $, $\frac i%
2\gamma _\mu \gamma _\nu \int F^{\mu \nu }\partial _\lambda { \psi \,}%
dx^\lambda $ cannot be judged reasonably, so these terms cannot be
thrown off and may show their significance in some situations when
the wavelength of fermions is sufficiently long. For instance,
$\tilde F^{\mu \nu }$ is an operator, and it renders the results
of $\frac{i\,}2\gamma _\mu \gamma _\nu \tilde F^{\mu \nu }{ \psi
}$ dependent on the variation of the wave function ${ \psi }$.
Accordingly this sort of terms is possibly
relevant to the nonlocal effect of Aharonov-Bohm type [7]. The terms $\frac{%
i\,}2\gamma _\mu \gamma _\nu \int \tilde F^{\mu \nu }\partial _\lambda { %
\psi \,}dx^\lambda $, $\frac i2\gamma _\mu \gamma _\nu \int F^{\mu
\nu }\partial _\lambda { \psi \,}dx^\lambda $ would behave in
similar ways.
But the term $i\,\int A_\nu \partial _\lambda \partial ^\nu { \psi \,}%
dx^\lambda $ , without the factor $\gamma _\mu \gamma _\nu $ before it, is
the exceptional case as discussed in the following.

Comparing the Eq. (8.12) with Eq. (8.6), it is noted that there
lacks a mass term $-m^2{ \psi }$ in the right hand of Eq. (8.12).
There are two ways to remedy this flaw. First, one can directly
add a mass term to the right hand of equation (8.9), then the
equation (8.2) would have a nonzero term in its right hand too.
That obviously violates our original hypothesis that motion
equation is just the geodetic line. Another way out is to accept
the term $i\,\int A_\nu \partial _\lambda \partial ^\nu { \psi
\,}dx^\lambda $ as the mass term. The reason is that according to
the Klein-Gordon equation of the next section the double
derivatives on potential $A_\nu $ would induce a mass factor and
the integration $\int A_\nu { \,}dx^\lambda $ would induce a phase
factor depending on closed paths (loops) of fermions [7], thus the
total effect of the term $\int A_\nu \partial _\lambda
\partial ^\nu { \psi \,}dx^\lambda $ seems equal to a mass term for
motion equation of fermions. If the paths are not closed, then the mass of
fermions would become dependent on potential $A_\nu $. The claims of this
paragraph are only qualitative, further confirmation is necessary. The
relevant works are in process.

Apart from the above qualitative display, all of the four nontrivial terms, $%
\frac{i\,}2\gamma _\mu \gamma _\nu \tilde F^{\mu \nu }{ \psi }$, $%
i\,\int A_\nu \partial _\lambda \partial ^\nu { \psi \,}dx^\lambda $, $%
\frac{i\,}2\gamma _\mu \gamma _\nu \int \tilde F^{\mu \nu
}\partial _\lambda { \psi \,}dx^\lambda $ and $\frac i2\gamma _\mu
\gamma _\nu \int F^{\mu \nu }\partial _\lambda { \psi
\,}dx^\lambda $, are surely nonlocal since they obviously ruin the
local conservation law $\partial _\mu j^\mu =0$ [where $j^\mu ={
\psi }\gamma ^\mu { \psi }$], which can be directly derived from
the conventional Dirac equation (8.5).

\subsection{Brief summary for this section}

If the approximation $(A_{\bar \rho \gamma })_{4\times 4}=\gamma _0+\gamma
_0\gamma ^\mu A_\mu $ is not used here, the replacement $d\rightarrow $ $%
\gamma _\mu \partial ^\mu $ would not be reasonable. In general case the
matrix $(A_{\bar \rho \gamma })_{4\times 4}$ should be expanded to all
possibilities of $\gamma -$matrices (the total number of matrices is 16)
without respecting only the vector form of interaction. But we think there
leaves little arbitrariness to extend the theory to include more sorts of
interaction. If surely the approximation $(A_{\bar \rho \gamma })_{4\times
4}=\gamma _0+\gamma _0\gamma ^\mu A_\mu $ holds for fundamental
interactions, then accordingly we would extend the replacement $d\rightarrow
$ $\gamma _\mu \partial ^\mu $ to $d\rightarrow $ $\lambda ^a\gamma _\mu
\partial _a^\mu $, $\lambda ^a$--Gellmann matrices, when we are concerned
about the colour interaction between quarks. That will cause more
integrations and other intricate terms to enter into Eq. (8.12). We will not
detail that issue here.

In conventional quantum field theory (CQFT), we only regard the part $F^{\mu
\nu }$ being responsible for the physical process. However, in view of the
nonlocal (curving) effects, it has been stated in Ref. [8] that the $F^{\mu
\nu }$ is not complete in describing all physics. Even under our rough
approximation, it can be seen in Eq. (8.12) that more terms are present than
required. It includes not only the normal terms $F^{\mu \nu }$ but also some
other terms on account of nonlocal effect.

Moreover, the definition of the connection and thus the physical meaning of $%
A^\mu $ now are not the same as those in CQFT, since in CQFT, $A^\mu $
(potential) is assumed as connection and $F^{\mu \nu }$ as the curvature
tensor. But here the $A^\mu $ is viewed only as components of metric tensor
and $F^{\mu \nu }$ as connection (force). Correspondingly the dimension of
motion equation here is also different from that of Schr\"odinger equation,
the former is a force, and the latter a potential. A discrepancy of energy
dimension appears under natural unit.

\section{Field equation for Bosons}

\subsection{The rule of calculating the derivative $\frac \partial {\partial
{ \bar z}^j}\frac \partial {\partial { z}^k}$}

The differential problem in fact has arisen in calculating the terms $dd{ %
\psi }^\beta $ and $\Gamma _{a\gamma }^\beta d{ \psi }^\alpha d{ \psi }%
^\gamma $. As a hypothesis, we have required the wave function to change
with respect to space-time variables $x_\mu $, with which the terms in Dirac
equation for electrodynamics are restored. However, in fact, since we are
discussing the property in complex space, the differential $dd{ \psi }%
^\beta $ should change with respect to complex variables as showed in Eq.
(8.3), i.e. $dd{ \psi }^{\prime \beta }=\frac{\partial ^2{ \psi }%
^{\prime \beta }}{\partial { \psi }^\alpha \partial { \bar \psi }%
^\gamma }d{ \psi }^\alpha \wedge d{ \bar \psi }^\gamma $. Making
every coefficient of $d{ \psi }^\rho \wedge d{ \bar \psi }^\sigma
$ vanish, then Eq.(8.2) yields
\begin{equation}
\frac{\partial ^2{ \psi }^{\prime \beta }}{\partial { \psi }^\rho
\partial { \bar \psi }^\sigma }+\Gamma _{a\gamma }^\beta \frac{\partial
{ \psi }^{\prime \alpha }}{\partial { \psi }^\rho }\frac{\partial { %
\psi }^{\prime \gamma }}{\partial { \bar \psi }^\sigma }=0
\tag{9.1}
\end{equation}
This equation should be thoroughly respected by the electron as an equal
level observer to another observed electron.

Although it is impossible for the electron to release all information it
carries by projecting its complex space to our space-time, ${ \psi }%
^\alpha \rightarrow x^\mu $, we have to do so to get some physical
observerbles in space-time. Just for the physical reason and for
consistence, in this section we directly make the replacements of the
differential forms $dd$ and $d{ A}d{ B}$, or the equivalent forms $%
\frac{\partial ^2}{\partial { \psi }^\rho \partial { \bar \psi }%
^\sigma }$ and $\frac{\partial { A}}{\partial { \psi }^\rho }\frac{%
\partial { B}}{\partial { \bar \psi }^\sigma }$, as follows:
\begin{equation}
dd\text{ or }\frac{\partial ^2}{\partial { \psi }^\rho \partial {
\bar \psi }^\sigma }\longrightarrow\Box  \tag{9.2a}
\end{equation}
\begin{equation}
d{ A}d{ B}\text{ or }\frac{\partial { A}}{\partial { \psi }^\rho
}\frac{\partial { B}}{\partial { \bar \psi }^\sigma }
\longrightarrow \frac{\partial { A}}{\partial x^\mu
}\frac{\partial { B}}{\partial x_\mu }  \tag{9.2b}
\end{equation}

\subsection{The antisymmetry of Ricci tensor and the operator $\Box $}

This subsection is devoted to explaining a subtle but important aspect in
calculating the curvature tensor. It can be seen from the definition of
Eq.(4.24) that the curvature tensor $R_{\alpha jk}^\beta $ is explicitly
antisymmetric with respect to the indices $j$ and $k$. However, the
antisymmetry seems lost from the expression (5.14). Now let us impose that
antisymmetry on the interpretation of Eq.(5.14) and see what will be the
result. Substitute the explicit form of connection $\Gamma _{\;\beta
}^\alpha $ into Eq.(5.14), the components of curvature become explicit
functions with respect to metric tensor:
\begin{eqnarray}
R_{\beta \bar jk}^\alpha &=&\frac \partial {\partial { \bar z}^j}%
(A^{\alpha \bar \gamma }\frac{\partial A_{\bar \gamma \beta }}{\partial { %
z}^k})\text{,}\;R_{\beta j\bar k}^\alpha =-\frac \partial
{\partial {
\bar z}^k}(A^{\alpha \bar \gamma }\frac{\partial A_{\bar \gamma \beta }}{%
\partial { z}^j})\text{,}  \nonumber \\
R_{\bar \beta \bar jk}^{\bar \alpha } &=&\frac \partial {\partial { \bar z%
}^j}(A^{\bar \alpha \gamma }\frac{\partial A_{\gamma \bar \beta }}{\partial
{ z}^k})\text{,}\;R_{\bar \beta j\bar k}^{\bar \alpha }=-\frac \partial {%
\partial { \bar z}^k}(A^{\bar \alpha \gamma }\frac{\partial A_{\gamma
\bar \beta }}{\partial { z}^j})\text{.}      {(9.3)} \nonumber
\end{eqnarray}
We only calculate the first one as an example, the others can be obtained by
properly changing the indices. Impose the antisymmetric property on the
first equation of Eq.(9.3)
\begin{equation}
\frac \partial {\partial { \bar z}^j}(A^{\alpha \bar \gamma }\frac{%
\partial A_{\bar \gamma \beta }}{\partial { z}^k})=-\frac \partial {%
\partial { z}^k}(A^{\alpha \bar \gamma }\frac{\partial A_{\bar \gamma
\beta }}{\partial { \bar z}^j})\text{ ,}  \tag{9.4}
\end{equation}
then the left hand and the right hand can be changed respectively to
\begin{equation}
left=-\frac \partial {\partial { \bar z}^j}[\frac{\partial
A^{\alpha \bar \gamma }}{\partial { z}^k}A_{\bar \gamma \beta
}]=-\frac{\partial A^{\alpha \bar \gamma }}{\partial {
z}^k}\frac{\partial A_{\bar \gamma
\beta }}{\partial { \bar z}^j}-A_{\bar \gamma \beta }\frac \partial {%
\partial { \bar z}^j}\frac \partial {\partial { z}^k}A^{\alpha \bar
\gamma }\text{ ,}  \tag{9.5a}
\end{equation}
\begin{equation}
right=-\frac{\partial A^{\alpha \bar \gamma }}{\partial { z}^k}\frac{%
\partial A_{\bar \gamma \beta }}{\partial { \bar z}^j}-A^{\alpha \bar
\gamma }\frac \partial {\partial { z}^k}\frac \partial {\partial {
\bar z}^j}A_{\bar \gamma \beta }\text{ ,} \tag{9.5b}
\end{equation}
the equality of the two sides directly gives
\begin{equation}
A^{\alpha \bar \gamma }\frac \partial {\partial { z}^k}\frac \partial {%
\partial { \bar z}^j}A_{\bar \gamma \beta }=A_{\bar \gamma \beta }\frac
\partial {\partial { \bar z}^j}\frac \partial {\partial { z}^k}%
A^{\alpha \bar \gamma }\text{ ,}  \tag{9.6a}
\end{equation}
i.e.
\begin{equation}
A^{\alpha \bar \gamma }\partial \bar \partial A_{\bar \gamma \beta
}=A_{\bar \gamma \beta }\bar \partial \partial A^{\alpha \bar
\gamma }\text{ .} \tag{9.6b}
\end{equation}
The relation seems trivial in this form, but making the replacement of (9.2)
results in
\begin{equation}
A^{\alpha \bar \gamma }\Box A_{\bar \gamma \beta }=A_{\bar \gamma
\beta }\Box A^{\alpha \bar \gamma }\text{ ,}  \tag{9.7}
\end{equation}
which is a reasonable result telling that the operator $\Box $ is Hermitian.
So now we can insist on the antisymmetry in Ricci tensor without worrying
about any unexpected contradiction.

As a byproduct of applying the above argument to Ricci tensor (5.18), a rule
is gained that for any product of two derivatives like $\frac \partial {%
\partial { \bar z}^j}\frac \partial {\partial { \bar z}^k}A$, $A$
arbitrary, the result should be antisymmetric if permuting $\frac \partial {%
\partial { \bar z}^j}$ and $\frac \partial {\partial { \bar z}^k}$, $%
\frac \partial {\partial { \bar z}^j}\frac \partial {\partial { \bar z}%
^k}A=-\frac \partial {\partial { \bar z}^k}\frac \partial
{\partial { \bar z}^j}A$.

\subsection{Field Equation}

Now let's introduce the field equation for bosons, e.g. photons in
Electrodynamics. For a boson field without any source, we are concerned
about the case when the interaction is absent, i.e., the connection (force)
of complex space is trivial, and thus all the components of curvature tensor
$R_{\beta jk}^\alpha $ vanish. In that case the boson field satisfies the
equation $R_{\beta jk}^\alpha =0$. The equation will not change under
certain transformations in question (e. g. $GL(4,C)$ group for photons)
since $R_{\beta jk}^\alpha $ is a tensor. That is to say, the field equation
will not change with the appearance of interaction. But as a general
equation for a boson field the condition $R_{\beta jk}^\alpha =0$ seems too
strict, so we introduce a weaker constraint by merely demanding that Ricci
tensors vanish,
\begin{equation}
R_{\alpha \beta }=0\text{ .}  \tag{9.8a}
\end{equation}
That not only holds for free fields, but also is expected to hold while a
source appears at a point infinitely far away. As in GR, we can add the
source term to the right hand of equation, whose physical meaning will be
clarified later,
\begin{equation}
R_{\alpha \beta }=\kappa S_{\alpha \beta }\text{ .}  \tag{9.8b}
\end{equation}
The coefficient $\kappa $ can be determined by comparing it with the
Klein-Gordon equation. Next we will study what can be derived from this
field equation, and what are its differences with the Klein-Gordon equation
as well.

\subsection{The Approximation of Metric Matrix With the Scalar Part Only}

We had better resolve the field equation (9.8) precisely before discussing
the field property. But here we expect to understand the field properties by
substituting some good approximations into the field equation.

As argued in section VI, in the absence of interaction the metric matrix is
of the form
\begin{equation}
\gamma _0=\left(
\begin{array}{cccc}
1 & 0 & 0 & 0 \\
0 & 1 & 0 & 0 \\
0 & 0 & -1 & 0 \\
0 & 0 & 0 & -1
\end{array}
\right) \text{ ,}  \tag{9.9}
\end{equation}
and after a period of interaction, according to the perturbative theory (in
this whole paper we respect the perturbative results of CQFT), the matrix
evolves into
\begin{equation}
\gamma _0+\gamma _0\gamma ^\mu A_\mu =\left(
\begin{array}{cc}
1+A_0 & -\vec \sigma \cdot \vec A \\
-\vec \sigma \cdot \vec A & -1+A_0
\end{array}
\right) \text{ .}  \tag{9.10}
\end{equation}
If only the electronic part is present, Eq. (9.10) gives
\begin{equation}
A_{\alpha \beta }\simeq \left(
\begin{array}{cccc}
1+A_0 & 0 & 0 & 0 \\
0 & 1+A_0 & 0 & 0 \\
0 & 0 & -1+A_0 & 0 \\
0 & 0 & 0 & -1+A_0
\end{array}
\right) \text{ .}  \tag{9.11}
\end{equation}
Moreover, if we choose the large components approximation, which is
validated by QED, then the form of $A_{\alpha \beta }$ yields
\begin{equation}
A_{\alpha \beta }\simeq \left(
\begin{array}{cccc}
1+A_0 & 0 & 0 & 0 \\
0 & 1+A_0 & 0 & 0 \\
0 & 0 & -1 & 0 \\
0 & 0 & 0 & -1
\end{array}
\right) \text{ .}  \tag{9.12}
\end{equation}

Now we briefly review the reasonability of the above approximation. The
large components and the small components appear in solution of Dirac
equation when both the electric part (scalar potential) and magnetic part
(vector part) of the boson field are small and the kinetic energy of fermion
is very low: In general, the Dirac equation can be written [2],
\begin{equation}
H{ \psi }=(m+W){ \psi }=\vec \alpha \cdot (\vec P-\vec A){ \psi }%
+\beta \,m{ \psi }+e\phi { \psi }\text{ ,}  \tag{9.13}
\end{equation}
where $W$ is kinetic energy, $\phi $ the scalar potential and
$\vec A$ the vector potential. Divide the four components ${ \psi
}$ into two parts composed respectively of the first two
components and the last two components,
\begin{equation}
{ \psi =}\left(
\begin{array}{c}
{ \psi }_1 \\
{ \psi }_2 \\
{ \psi }_3 \\
{ \psi }_4
\end{array}
\right) =\left(
\begin{array}{c}
{ \psi }_a \\
{ \psi }_b
\end{array}
\right) \text{,}  \tag{9.14}
\end{equation}
with this denotation the Dirac equation is readily reduced into two
equations,
\begin{eqnarray}
(m+W){ \psi }_a &=&\vec \sigma \cdot (\vec P-\vec A){ \psi }_b+m{ %
\psi }_a+e\phi { \psi }_a  \nonumber \\
(m+W){ \psi }_b &=&\vec \sigma \cdot (\vec P-\vec A){ \psi }_a-m{ %
\psi }_b+e\phi { \psi }_b\;\text{.}  {(9.15)} \nonumber
\end{eqnarray}
After carrying out ${ \psi }_b$ as ${ \psi }_b=(2m+W-\phi
)^{-1}\,\vec \sigma \cdot (\vec P-\vec A){ \psi }_a$, we see that
when $W $ , $\phi $
and $\vec A$ are all small, the two components ${ \psi }_a$ and ${ %
\psi }_b$ obey the relation
\begin{equation}
{ \psi }_b\sim \frac{\vec P}m{ \psi }_a\sim \frac \upsilon c{ \psi }%
_a\text{ .}  \tag{9.16}
\end{equation}
So in the large--component--approximation{\bf ,} the potential $A_0$ in the
last two diagonal elements in (9.11) can be ignored reasonably since it
contributes to ${ \psi }^\alpha A_{\alpha \bar \beta }{ \bar \psi }%
^\beta $ the terms two orders less than that of the first two diagonal
elements. Consequently the approximation (9.12) does hold.

Under this approximation, the Ricci tensor reads
\begin{equation}
R_{\alpha \beta }=\frac{\partial ^2\ln \det \left(
\begin{array}{cccc}
1+A_0 & 0 & 0 & 0 \\
0 & 1+A_0 & 0 & 0 \\
0 & 0 & -1 & 0 \\
0 & 0 & 0 & -1
\end{array}
\right) }{\partial { \psi }^\alpha \partial { \psi }^\beta }\text{
.}  \tag{9.17}
\end{equation}
Replacing the differential $\frac{\partial ^2}{\partial { \psi
}^\alpha
\partial { \psi }^\beta }$ following Eq. (9.2), and assuming that the
scalar field does not vary with time, then up to $(A^0)^3$ order the field
equation $R_{\alpha \beta }=0$ turns out to be
\begin{equation}
\vec \nabla ^2A^0=0  \tag{9.18}
\end{equation}
which is just the Poisson equation satisfied by the electric field in
vacuum. In the next subsection it will be shown that the approximation
without the electric field is also heuristic.

\subsection{The Approximation of Metric Matrix With the Vector Part Only and
a Conjecture on the Origin of Mass for Bosons}

Now let's turn to the approximation without electric field or source term,
i.e., the case that only radiation field exists. Computing the similar form
of Ricci tensor in (9.17) by using the metric form (9.10) and making $A_0=0$%
, we have

\begin{equation}
\Box \vec A^2=0\text{ .}  \tag{9.19}
\end{equation}
We will treat the equation explicitly to construct its relation to the
electric and magnetic energy forms $\vec E^2$ and $\vec B^2$, where $\vec E=-%
\frac{\partial \vec A}{\partial t}$ and $\vec B=\vec \nabla \times \vec A$.
The calculation of (9.19) includes two parts
\begin{equation}
\frac{\partial ^2}{\partial t^2}\vec A^2=2(\frac \partial {\partial t}\vec A%
)^2+2\vec A\cdot \frac{\partial ^2}{\partial t^2}\vec A\text{ ,}
\tag{9.20}
\end{equation}
and
\begin{eqnarray}
\vec \nabla ^2\vec A^2 &=&\vec \nabla \cdot (\vec \nabla \vec A^2)=\vec
\nabla \cdot (2\underline{\vec A}\cdot \vec \nabla \underline{\vec A})
\nonumber \\
\ &=&2(\vec \nabla \underline{\vec A}):(\vec \nabla \underline{\vec A})+2%
\underline{\vec A}\cdot \vec \nabla ^2\underline{\vec A}\text{ ,}
{(9.21)}  \nonumber
\end{eqnarray}
where the underlines denote the inner products between the vectors. And the
sign $:$ is the tensor product for dyads, for example, $\vec p\cdot \vec
\sigma _1\vec k\cdot \vec \sigma _2=$ $\vec p\vec k:\vec \sigma _1\vec \sigma
_2$. If an inner product is emphasized by underlines then the regular
product order between dyads is not followed, e.g. $\vec p\underline{\vec k}:%
\underline{\vec \sigma _1}\vec \sigma _2=\vec k\cdot \vec \sigma _1\,\vec p%
\cdot \vec \sigma _2$. Usually we only use the underlines while the gradient
operator $\vec \nabla $ appears. Now let's express the magnetic energy $\vec
B^2$ in terms of vector potential
\begin{eqnarray}
\vec B^2 &=&(\vec \nabla \times \vec A)^2=(\vec \nabla \times \vec A)\cdot (%
\vec \nabla \times \vec A)  \nonumber \\
\ &=&\vec \nabla \cdot (\vec A\times (\vec \nabla \times \vec A))+\vec A%
\cdot (\vec \nabla \times (\vec \nabla \times \vec A))\text{ .}
{(9.22)}  \nonumber
\end{eqnarray}
To further simplify the Eq. (9.22), the following relations are useful,
\begin{eqnarray}
2(\vec A\times (\vec \nabla \times \vec A)) &=&\vec \nabla (\vec A\cdot \vec
A)-2(\vec A\cdot \vec \nabla )\vec A  \nonumber \\
\ &=&2\underline{\vec A}\cdot (\vec \nabla \underline{\vec
A})-2(\vec A\cdot \vec \nabla )\vec A\text{ ,}  {(9.23a)}
\nonumber
\end{eqnarray}
\begin{eqnarray}
\vec \nabla \cdot (\vec A\times (\vec \nabla \times \vec A)) &=&\vec \nabla
\cdot [\underline{\vec A}\cdot (\vec \nabla \underline{\vec A})-(\vec A\cdot
\vec \nabla )\vec A]  \nonumber \\
\ &=&(\vec \nabla \underline{\vec A}):(\vec \nabla \underline{\vec A})+\vec A%
\cdot \vec \nabla ^2\vec A-\vec \nabla \underline{\vec
A}:\underline{\vec \nabla }\vec A-(\vec A\cdot \vec \nabla )(\vec
\nabla \cdot \vec A)\text{ ,}        {(9.23b)} \nonumber
\end{eqnarray}
\begin{equation}
\vec A\cdot \vec \nabla \times (\vec \nabla \times \vec A)=\vec
A\cdot [\vec \nabla (\vec \nabla \cdot \vec A)-\vec \nabla ^2\vec
A]=(\vec A\cdot \vec \nabla )(\vec \nabla \cdot \vec A)-\vec
A\cdot \vec \nabla ^2\vec A\text{ .} \tag{9.23c}
\end{equation}
Applying relations in (9.23) to (9.22), we obtain
\begin{equation}
\vec B^2=(\vec \nabla \underline{\vec A}):(\vec \nabla \underline{\vec A})-%
\vec \nabla \underline{\vec A}:\underline{\vec \nabla }\vec
A=(\vec \nabla \underline{\vec A}):(\vec \nabla \underline{\vec
A})\text{ .}  \tag{9.24}
\end{equation}
The last step holds for transverse fields in the Coulomb gauge, $\vec \nabla
\cdot \vec A=\underline{\vec A}\cdot \underline{\vec \nabla }\vec A=0$. Now
the Eq. (9.20) and Eq. (9.21) can be simplified to
\begin{equation}
\frac{\partial ^2}{\partial t^2}\vec A^2 =2\vec E^2+2\vec A\cdot \frac{%
\partial ^2}{\partial t^2}\vec A\text{ ,}  \tag{9.25a}
\end{equation}
\begin{equation}
-\vec \nabla ^2\vec A^2 =-2\vec B^2-2\vec A\cdot \vec \nabla
^2\vec A\text{ .}  \tag{9.25b}
\end{equation}
Combine them into (9.19)
\begin{equation}
\Box \vec A^2=2(\vec E^2-\vec B^2)+2\vec A\cdot \Box \vec
A=0\text{.} \tag{9.26}
\end{equation}
Consequently, the following equation holds,
\begin{equation}
\vec A\cdot \Box \vec A=\vec B^2-\vec E^2\stackrel{d}{=}-m^2\vec
A^2\text{.} \tag{9.27}
\end{equation}

Heuristically, the Eq. (9.27) suggests if the inequality
\begin{equation}
\vec E^2-\vec B^2\neq 0\text{ }  \tag{9.28}
\end{equation}
holds, then the equation for field $\vec A\ $can automatically gain a mass
term
\begin{equation}
\vec A\cdot (\Box +m^2)\vec A=0\text{ ,}  \tag{9.29}
\end{equation}
where the Klein-Gordon equation takes shape. For Electrodynamics, since $%
\vec E^2-\vec B^2$ coincidentally and accurately vanishes, the net mass of
photon is zero. In QED, the Lagrangian density for photon field is
\begin{equation}
{ L}=-\frac 14F_{\mu \nu }(x)F^{\mu \nu }(x)\text{ ,} \tag{9.30}
\end{equation}
the field tensor $F_{\mu \nu }(x)$ is defined in the preceding
section. In terms of $\vec E$ and $\vec B$, the Lagrangian can be
transformed to
\begin{equation}
{ L}=-\frac 14F_{\mu \nu }(x)F^{\mu \nu }(x)=-\frac 12(\vec E^2-\vec B%
^2)=-\frac 12\,m^2\vec A^2\text{ .}  \tag{9.31}
\end{equation}
Coincident with Lagrange undetermined multiplier method, now it is natural
to extend the Lagrangian of QED by adding a mass term
\begin{equation}
\ { L}=-\frac 14F_{\mu \nu }(x)F^{\mu \nu }(x)+\frac 12\,m^2A_\mu
A^\mu \text{ ,}  \tag{9.30}
\end{equation}
from which the Klein-Gordon equation for massive bosons can be obtained by
using variational method. In the frame of this paper we don't respect gauge
invariance of the Lagrangian. But the variational method would always be
valid from mathematical viewpoint.

Now let's clarify the meaning of source term in Eq. (9.8b). In quantum field
theory the Klein-Gordon equation with source term is $\Box \vec A=\vec j$,
hence according to Eq. (9.29)
\begin{equation}
\vec A\cdot (\Box +m^2)\vec A=\vec A\cdot \vec j=\vec A\cdot {\it \bar \psi }%
\vec \gamma {\it \psi }\text{ ,}  \tag{9.31}
\end{equation}
the right hand is just the interaction Hamiltonian. Remembering that we have
in Eq. (9.2) replaced the derivatives with respect to ${\it \psi }$ by the
derivatives with respect to space-time, and noticing the form $\frac \partial
{\partial {\it \psi }}=\frac \partial {\partial x}\frac{\partial x}{\partial
{\it \psi }}$, the above equation and Eq. (9.17) suggest that the source
term corresponding to Eq. (9.8) should be of a double integrals with $\not
A=A_\mu \gamma ^\mu $ as integrand and $(\gamma _\nu dx^\nu )^2$ as its
infinitesimal integral measurement. We define $\tilde A_{\alpha \beta }=\int
\int A_\mu \gamma ^\mu (\gamma _\nu dx^\nu )^2$, then
\begin{equation}
R_{\alpha \beta }=\lambda \tilde A_{\alpha \beta }\text{,}
\tag{9.32}
\end{equation}
here the $\tilde A_{\alpha \beta }$ refers to a source field induced by
fermions, though it looks formally like from $A_\mu $. The term $\vec E^2-%
\vec B^2$ should not be associated with $\tilde A_{\alpha \beta }$ since it
is derived from the equation for free boson field.

\subsection{The general forms of field equation in space-time}

In above subsections the discussions are applicable in fact only to $U(1)$
case, in which $A_\mu $ has no intrinsic degrees of freedom except spatial
polarization. As stated in the previous section, if other intrinsic degrees
are involved [e. g. see (7.3), $A_\mu \rightarrow A_{\alpha \beta }^{ac}$],
the replacement $d\rightarrow $ $\gamma _\mu \partial ^\mu $ will not hold
any longer. Consequently, the replacement of Eq. (9.2) becomes abated.
Instead we should employ the replacement $d\rightarrow $ $\tau ^a\gamma _\mu
\partial _a^\mu $, $\tau ^a$ being generators of a Unitary group. When the
group is $SU(2)$, whose representation can use Pauli matrices, the
replacement performing on double differential $dd$ will induce the similar
form to $U(1)$ case, $dd\rightarrow \Box _1^2+\Box _2^2+\Box _3^2$. In this
case the field equation in space-time would be
\begin{equation}
(\Box _1^2+\Box _2^2+\Box _3^2)\det (A_{\alpha \beta
}^{ac})=0\text{ .} \tag{9.33}
\end{equation}
However, in $SU(3)$ case, the Gellmann matrices $\{\lambda ^a,a=1,\cdots
,8\} $ have not the property of $(\lambda ^ak_a)(\lambda
^bk_b)=k_1^2+k_2^2+\cdots +k_8^2$, so that the cross-terms don't vanish.
Therefore, in $SU(3)$ case, the field equation in space-time can written
only as
\begin{equation}
(\lambda ^a\gamma _\mu \partial _a^\mu )(\lambda ^b\gamma _\nu
\partial _b^\nu )\det (A_{\alpha \beta }^{ac})=\lambda ^a\lambda
^b\gamma _\mu \gamma _\nu \partial _a^\mu \partial _b^\nu \det
(A_{\alpha \beta }^{ac})=0\text{ .} \tag{9.34}
\end{equation}

\section{The Bianchi Identity and Remarks on Conservation Law}

Make exterior differential to both sides of Eq. (4.23), one gets the
following Bianchi identity
\begin{equation}
d\Omega =\Omega \wedge \Gamma -\Gamma \wedge \Omega \text{ ,}
\tag{10.1}
\end{equation}
and its matrix form is
\begin{equation}
d\Omega _{\;k}^l=\Omega _{\;h}^l\wedge \Gamma _{\;k}^h-\Gamma
_{\;h}^l\wedge \Omega _{\;k}^h\text{.}  \tag{10.2}
\end{equation}
Write it out in components form
\begin{eqnarray}
\frac{\partial R_{bjk}^a}{\partial { z}^m}d{ z}^m\wedge d{ z}%
^j\wedge d{ z}^k &=&R_{cij}^ad{ z}^i\wedge d{ z}^j\wedge \Gamma
_{mb}^cd{ z}^m-\Gamma _{mc}^ad{ z}^m\wedge R_{bij}^cd{ z}^i\wedge d%
{ z}^j  \nonumber \\
\ &=&(R_{cmi}^a\Gamma _{jb}^c-\Gamma _{mc}^aR_{bij}^c)d{ z}^m\wedge d{ %
z}^i\wedge d{ z}^j\;\text{.}   {(10.3)} \nonumber
\end{eqnarray}
On the other hand, the covariant differential of $R_{bij}^a$ can be obtained
by definition
\begin{equation}
R_{bij;m}^a=\frac{\partial R_{bij}^a}{\partial { z}^m}+\Gamma
_{mc}^aR_{bij}^c-\Gamma _{mb}^cR_{cij}^a-\Gamma
_{mi}^cR_{bcj}^a-\Gamma _{mj}^cR_{bic}^a\;\text{.}  \tag{10.4}
\end{equation}
Combining the above two, one gets
\begin{equation}
R_{bij;m}^ad{ z}^m\wedge d{ z}^i\wedge d{ z}^j=-(\Gamma
_{mi}^cR_{bcj}^a+\Gamma _{mj}^cR_{bic}^a)d{ z}^m\wedge d{ z}^i\wedge d%
{ z}^j\;\text{.}  \tag{10.5}
\end{equation}
In the case that torsion is absent, the right side is equal to zero, hence
\begin{equation}
R_{b[ij;m]}^a=0\;\text{.}  \tag{10.6}
\end{equation}
The bracket $[ij;m]$ means the left side also includes the terms
with the indices of cyclic permutation. Substituting the nonzero
components in Eq.(5.14) one by one to the above equation, for
instance substituting the (5.14a), it becomes
\begin{eqnarray}
R_{\beta [\bar jk;m]}^\alpha &=&R_{\beta \bar jk;m}^\alpha +R_{\beta m\bar j%
;k}^\alpha +R_{\beta km;\bar j}^\alpha  \nonumber \\
\ &=&R_{\beta \bar jk;m}^\alpha +R_{\beta m\bar j;k}^\alpha \stackrel{\text{%
contracting }\alpha \text{ and }\beta }{=}R_{\bar jk;m}+R_{m\bar j;k}=0\;%
\text{.}  {(10.7)} \nonumber
\end{eqnarray}
The other three forms of Eq.(5.14) will give rise to the same form of the
Ricci identity $R_{\bar jk;m}+R_{m\bar j;k}=0$. It has been proved in Sec.V
that the metric is invariant under covariant differential, $A_{\alpha \beta
;j}=D_jA_{\alpha \beta }=0$. So with the definition of scalar curvature $%
R=A^{\alpha \bar \beta }R_{\alpha \bar \beta }$, the Ricci identity
possesses the following form
\begin{equation}
(A^{m\bar j}R_{m\bar j})_{;k}+(A^{m\bar j}R_{\bar
jk})_{;m}=0\;\text{,} \tag{10.8}
\end{equation}
i.e.,
\begin{equation}
R_{;k}-R_{\;\;k;m}^m=0\;\text{.}  \tag{10.9}
\end{equation}
Applying the relation $A_{\alpha \beta ;j}=0$ to above equation, with $A^{%
\bar \rho k}R_{\;\;k}^m=R^{\bar \rho m}$, the Eq. (10.9) can be changed to
\begin{equation}
(R^{\bar \rho m}-A^{\bar \rho m}R)_{;m}=0\;\text{.}  \tag{10.10}
\end{equation}

Now the field equation can be rewritten by regarding the Bianchi identity,
\begin{equation}
R^{\bar \rho m}-A^{\bar \rho m}R=0\text{ ,}  \tag{10.11}
\end{equation}
its equivalence to Eq. (9.8) can be demonstrated as follows: contracting the
indices of Eq. (10.11) results in $R-4R=0$, then substituting $R=0$ back to
Eq. (10.11), finally the Eq. (9.8) is restored. Furthermore, the above
equation can be extended to include source term
\begin{equation}
R^{\bar \rho m}-A^{\bar \rho m}R=\kappa S^{\bar \rho m}\;\text{.}
\tag{10.12}
\end{equation}

As mentioned in the last section, the approximation (9.10) is good for
perturbative theories. In such case, the metric reads [9]
\begin{equation}
A_{\alpha \beta }=\eta _{\alpha \beta }+h_{\alpha \beta }\text{ ,}
\tag{10.13}
\end{equation}
where the $h_{\alpha \beta }$ is required to be small comparing
with $\eta _{\alpha \beta }$, whose nonvanishing elements are only
those diagonal ones, being $1,1,-1,-1$. Here the linear metric
$\eta _{\alpha \beta }$ is used to
raise or lower the indices of metric and curvature of linear part, e.g. $%
\eta ^{\bar \alpha \beta }h_{\beta \bar \gamma }=h_{\;\bar \gamma }^{\bar
\alpha }$. Only in this sense, the following defined $R_{\bar \alpha \beta
}^{{\bf L}}$ is different from the normal definition of the Ricci tensor $R_{%
\bar \alpha \beta }$, though they have almost the same form. The linear part
of the Ricci tensor is
\begin{equation}
R_{\bar \alpha \beta }^{{\bf L}}=\frac{\partial ^2h_{\;\lambda }^\lambda }{%
\partial { \bar z}^\alpha { z}^\beta }\;\text{.}  \tag{10.14}
\end{equation}
The rest part then is
\begin{equation}
R_{\bar \alpha \beta }-R_{\bar \alpha \beta }^{{\bf L}}=\frac \partial {%
\partial { \bar z}^\alpha }(h^{\bar \delta \gamma }\frac{\partial
h_{\gamma \bar \delta }}{\partial { z}^\beta })\;\text{.}
\tag{10.15}
\end{equation}
Similar to the treatment in General{\bf \ }Relativity [{\bf 7}], we separate
the left hand of the field Eq. (10.12) into two parts $R_{\bar \rho m}-A_{%
\bar \rho m}R=R_{\bar \rho m}^{{\bf L}}-\eta _{\bar \rho m}R^{{\bf L}%
}-\kappa \Pi _{\bar \rho m}$, then Eq. (10.12) can be written
\begin{equation}
R_{\bar \rho m}^{{\bf L}}-\eta _{\bar \rho m}R^{{\bf L}}=\kappa
(\Pi _{\bar \rho m}+S_{\bar \rho m})\text{ .}  \tag{10.16}
\end{equation}
We can prove that the differential performed on the left hand of Eq. (10.16)
gives zero
\begin{equation}
\frac \partial {\partial { z}^m}(R_{\bar \rho m}^{{\bf L}}-\eta
_{\bar \rho m}R^{{\bf L}})=0\;\text{.}  \tag{10.17}
\end{equation}
The first term is zero since the antisymmetric tensor $\frac \partial {%
\partial { z}^m}R^{\mu \bar \nu }($Indices $m,\bar \nu $ are antisymmetric%
$)$ is multiplied by symmetric tensor $\eta _{m\bar \nu }$, i.e., $R_{\bar
\rho m}^{{\bf L}}=\eta _{\bar \rho \mu }\eta _{m\bar \nu }R^{\mu \bar \nu }$%
; and the second term is trivial since the symmetric tensors $\eta _{n\bar
\nu }$,$\eta _{\bar \sigma \mu }$ and $\eta ^{\bar \sigma n}$ are
accompanied with the antisymmetric tensor $R^{\mu \bar \nu }$, so $\eta ^{%
\bar \sigma n}\eta _{\bar \sigma \mu }\eta _{n\bar \nu }R^{\mu \bar \nu }=R^{%
{\bf L}}$ is directly zero before differential is performed.

From (10.10), (10.12) we conclude that the source term $S_{\bar \rho m}$ is
conservative in covariant sense, $S_{\bar \rho m;m}=0$. Whereas from
(10.16), (10.17) we conclude that the sum $(\Pi _{\bar \rho m}+S_{\bar \rho
m})$ is conservative in sense of common differential, $\frac \partial {%
\partial { z}^m}(\Pi _{\bar \rho m}+S_{\bar \rho m})=0$. The two sorts of
conservation directly bring out three conclusions:

(I) As for the existence of the conserved equation
\begin{equation}
\frac \partial {\partial { z}^m}(\Pi _{\bar \rho m}+S_{\bar \rho
m})=0 \tag{10.18}
\end{equation}
it is natural to view the sum $\Pi _{\bar \rho m}+S_{\bar \rho m}$ as the
total energy-momentum tensor including both the fermion part $S_{\bar \rho
m} $ and the boson part $\Pi _{\bar \rho m}$, which suggests in a
complex--infinitesimal--local--region the energy and momentum may transfer
between $S_{\bar \rho m}$ and $\Pi _{\bar \rho m}$.

(II) Since the $R_{\bar \rho m}^{{\bf L}}$ will not be a tensor for the loss
of nonlinear terms, it can be inferred from (10.16) that the boson part $\Pi
_{\bar \rho m}$ does not satisfy $\Pi _{\bar \rho m;m}=0$ (this equation and
(10.18) can't be satisfied at the same time), and thus can't be a locally
conserved quantity. Partly for this reason the theory in this paper is {\bf %
nonlocal}.

(III) It seems that the aforementioned mass terms $\vec E^2-\vec B^2$ should
be included in the boson self-energy term $\Pi _{\bar \rho m}$, but the last
word on that requires more investigation. If that is the case, what about
the remaining parts in $\Pi _{\bar \rho m}$?

The conclusions of this section are obtained in the absence of torsion. If
torsion is present, then the forms of formulae will be more complicated.

\section{Physial Region}

\subsection{The Curved Space for Electromagnetic Field}

From now on we will work in the approximation of metric tensor shown in
(9.10):

\begin{equation}
(A_{\alpha \beta })_{4\times 4}=\gamma _0+\gamma _0\gamma ^\mu A_\mu =\left(
\begin{array}{cc}
1+A_0 & -\vec \sigma \cdot \vec A \\
-\vec \sigma \cdot \vec A & -1+A_0
\end{array}
\right) \text{,}  \tag{9.10}
\end{equation}
which is justified by the success of CQFT for perturbative interaction. The
metric, as stated in section VI, is
\begin{equation}
A(X,Y)=d{ \psi }^\alpha A_{\alpha \bar \beta }d{ \bar \psi }^\beta \;%
\text{,}  \tag{11.1}
\end{equation}
correspondingly the Ricci curvature for it is
\begin{equation}
R_{\bar \alpha \beta }=\frac{\partial ^2\ln A}{\partial { \bar
z}^\alpha { z}^\beta }\text{ .}  \tag{11.2}
\end{equation}
The present goal is to determine the physical region on the basis of the
above formulae. Substituting Eq. (9.10) to the Eq. (11.2), we find the main
quantity to be evaluated is the determinant of matrix $(A_{\alpha \bar \beta
})_{4\times 4}$. The determinant of (9.10) is
\begin{equation}
\mid A_{\bar \alpha \beta }\mid =1-A_0^2+\vec A^2\;\text{.}
\tag{11.3}
\end{equation}

Suggested by the horizon of black hole in GR, the horizon-like boundary of
the physical region may exist also for those fields with the same
theoretical frame as GR. As a tentative step, let's first examine the
physical region (or alternatively, the singularity) for electrons (whose
dynamics is governed by QED) to get some rules and then apply them to quarks
(The dynamics is mostly governed by QCD).

From the form of Eq.(11.2), it is noted that if
\begin{equation}
\mid A_{\bar \alpha \beta }\mid =1-A_0^2+\vec A^2=0\text{ ,}
\tag{11.4}
\end{equation}
then the logarithm function in Ricci tensor becomes divergent. The
solutions of the equation are defined as the singularities of
Ricci tensor. In the large--component--approximation and $\vec
A\sim 0$, we obtain
\begin{equation}
A_0\sim 1\text{ .}  \tag{11.5}
\end{equation}
For the electron in a hydrogen--atom, $A_0$ is the Coulomb form $\frac 1r$,
then Eq. (11.5) leads to $r\sim 1$. In the atomic (natural) unit, that means
the length of $r$ is the average radius of the ground state of the hydrogen
atom, which suggests that the zero singularity of logarithm function
possibly symbolizes one side of physical region. On the other hand, if $%
r\rightarrow \infty $, then $1-\frac 1{r^2}\rightarrow 1$, the electron
tends to be free (asymptotically free). In summary we achieve the following
conclusion held for electrons,
\begin{equation}
\mid A_{\bar \alpha \beta }\mid =1-A_0^2+\vec A^2=\{
\begin{array}{cc}
0 & \text{bound states} \\
1 & \text{ asymptotically--free states}
\end{array}
\tag{11.6}
\end{equation}
Now let's take this conclusion as a rule: even in other situations, $\mid A_{%
\bar \alpha \beta }\mid =0$ and $1$ represent the two sides of boundary of
{\bf physical region}, beyond which there are singularities [Fig. 2]. As for
a electron in hydrogen atom the singularity region ({\bf nonlocal region})
is the region within the range of proton's wavelength.

As for the bound-state side of physical region, even in
ElectroDynamics
[10], one may encounter alternatives in which a particular approximation of $%
\vec A=0$ seems unnecessary. For instance, that $1-A_0^2+\vec A^2=0$
resulting in $A_0^2=1+\vec A^2$ looks also reasonable. However, in a
perturbative theory there is no chance for any components of potential to
have a value larger than $1$. So in perturbative case, the only way to meet $%
1-A_0^2+\vec A^2=0$ is to make $\vec A=0$ and $A_0\rightarrow 1^{-}$. But
the requirement to the asymptotically--free side is loose, because in any
case the alternative solution $\mid A_0\mid \sim \mid \vec A\mid $
(additional to aforementioned $\vec A\rightarrow 0$, $A_0\rightarrow 0$) of $%
1-A_0^2+\vec A^2=1$ is by no means forbidden.

\subsection{Two Types of Curving in Colour Space}

In section VI, we have noted that the metric form is relevant to the
interaction vertex of CQFT. In QCD, we adopt the colour-spin-independent
form to express its interaction vertex. Apart from a coupling constant, the
vertex reads
\begin{equation}
\bar \Psi (x)\gamma _\mu A_a^\mu \lambda ^a\Psi (x)=\bar \Psi
(x)\gamma _\mu \lambda ^a\Psi (x)A_a^\mu \text{,}  \tag{11.7}
\end{equation}
in which $\lambda ^a$ ($a=1,\cdots ,8$) are Gell-mann matrices.
Correspondingly, the wave function for quarks can be also written in a
separable form
\begin{equation}
\Psi (x)={ \psi }(x)\left(
\begin{array}{c}
q^a(x) \\
q^b(x) \\
q^c(x)
\end{array}
\right) \text{,}  \tag{11.8}
\end{equation}
where ${ \psi }(x)$ is the familiar spinor part [being transformed
under the group $GL(4,\not C)$] and $\left(
\begin{array}{c}
q^a(x) \\
q^b(x) \\
q^c(x)
\end{array}
\right) $ the colour part [being transformed under the group $U(3,\not C)$].
In this respect, the complete form of metric for colour interaction can be
written
\begin{equation}
A(X,Y)=d{ \psi }^\alpha (x)q_a(x)A_{\alpha \bar \beta
}^{ab}q_b(x)d{ \bar \psi }^\beta (x)\text{,}  \tag{11.9}
\end{equation}
where the metric tensor $A_{\alpha \bar \beta }^{ab}$ may be further
decomposed into
\begin{equation}
A_{\alpha \bar \beta }^{ab}=A_{\alpha \bar \beta }A^{ab}\text{ or }%
(A_{\alpha \bar \beta }^{ab})_{12\times 12}=(A_{\alpha \bar \beta
})_{4\times 4}\otimes (A^{ab})_{3\times 3}\text{ .}  \tag{11.10}
\end{equation}

We have mentioned in section IV how the connection, curvature and the Ricci
tensor are derived from $A_{\alpha \bar \beta }$. Being multiplied by an
additional matrix element $A^{ab}$, the forms of the aforementioned geometry
quantities constructed from $A_{\alpha \bar \beta }$, particularly the Ricci
tensor, will not be changed since we can make the indices $a$, $b$ fixed
temporarily so that the factor $A^{ab}$ becomes a constant. The same
procedure is equally applicable to fixing the indices $\alpha $, $\bar \beta
$ and letting $(A^{ab})_{3\times 3}$ form the corresponding curvature and
Ricci tensor. Combining the two procedures we obtain the Ricci tensor for
the whole metric
\begin{equation}
R_{\bar \rho \sigma }=\frac{\partial ^2\ln \mid A_{\alpha \bar
\beta }^{ab}\mid }{\partial { \bar z}^\rho { z}^\sigma
}=3\frac{\partial ^2\ln \mid A_{\bar \alpha \beta }\mid }{\partial
{ \bar \psi }^\rho
\partial { \psi }^\sigma }+4\frac{\partial ^2\ln \mid A^{ab}\mid }{%
\partial \bar q^\rho q^\sigma }=R_S+R^C\text{ ,}  \tag{11.11}
\end{equation}
where $R_S$ is for spinor and $R^C$\ for colour{\bf .} In view of section
VII, we recognize that the metric tensor $A_{\bar \alpha \beta }$ is for the
hyperbolic space and tensor $A^{ab}$ is for ellipse colour space, i.e., $R_S$
and $R^C$ are subject to different geometries (different curvings) though
their forms are similar. Consequently, the field equation $R_{\bar \alpha
\beta }=0$ now becomes
\begin{equation}
R_S=-R^C\;\text{.}  \tag{11.12}
\end{equation}
For simplicity in form, we assume that the constants or other
variables before $\frac{\partial ^2\ln \mid A_{\bar \alpha \beta
}\mid }{\partial { \bar \psi }^\rho \partial { \psi }^\sigma }$
and $\frac{\partial ^2\ln
\mid A^{ab}\mid }{\partial \bar q^\rho q^\sigma }$ have been absorbed in $%
R_S $ and $R^C$.

The above equation provides us with a way to give rise to mass term
alternatively, or reversely to eliminate the mass term in Eq. (9.29). In
section-IX we have obtained the field equation in space-time $\vec A\cdot
\Box \vec A+(\vec E^2-\vec B^2)=0$, which can be derived from $R_S$ in
Eq.(11.12). On the other hand, $-R^C$ can contribute a nontrivial term to
the field equation in space-time, and roughly we denote such a term by $%
m_C^2\,\vec A^2$. So far the field equation for colour field is given by
\begin{equation}
\vec A\cdot \Box \vec A+(\vec E^2-\vec B^2)=m_C^2\,\vec A^2\text{
.} \tag{11.13}
\end{equation}
From this equation one recognizes that even when $\vec E^2-\vec B^2=0$ can a
mass term be provided by $m_C^2\,\vec A^2$, if only the dimension of boson
field is larger than $1$[$U(1)$]. Possibly the term $m_C^2\,\vec A^2$ can
offer a mechanism to give rise to mass for gluons. But the reverse
possibility exists simultaneously: if $\vec E^2-\vec B^2\neq 0$, the term $%
m_C^2\,\vec A^2$ may happen to eliminate the contribution from $\vec E^2-%
\vec B^2$---when their signs are the same. So at the present formula-level
it can't be asserted whether the gluon owns mass or not.

\subsection{Physical Region in Colour Space and two Understandings to Colour
Confinement}

From both experiments and the theory of QCD, it has been well known that the
colour interaction is asymptotically free, i.e. while the transferred
momenta of quarks are very large, the coupling constant of colour
interaction tends to zero and the interaction becomes a perturbative one. In
this case, following the condition $1-A_0^2+\vec A^2\rightarrow 1$ in
Eq.(11.6), we have $A_0\rightarrow 0$, $\vec A\rightarrow 0$ or $A_0\sim $ $%
\vec A$. The other side of boundary of quarks' physical region, where the
quarks are tightly confined, is assumed to correspond to the case of bound
electron in hydrogen atom. Now we begin to discuss this side of boundary of
physical region for colour interaction.

When two quarks are in their ground state (a low--momentum state), according
to Eq.(11.6) and the parallelism between confined quarks and bound electrons
[Fig. 2], the condition $\mid A_{\bar \alpha \beta }^{ab}\mid =0$ should
hold. The Eq.(11.11) means that either $\mid A_{\bar \alpha \beta }\mid =0$
or $\mid A^{ab}\mid =0$ is satisfied. Among others, $\mid A^{ab}\mid =0$
deserves more attention. Obviously it makes the rank of the colour matrix
decrease at least by one, from $SU(3)$ to $SU(2)$ or $U(1)$, which suggests
that quarks condense to hadrons, or possibly others, and thus no free quark
appears. This conclusion seems able to account for the confinement of
quarks, at least as a mechanism.

Next we examine the meaning of $\mid A_{\bar \alpha \beta }^{ab}\mid =0$ in
more details. Under our approximation (11.7), we have expressed the metric
matrix as $(A_{\bar \alpha \beta }^{ab})=(A_{\alpha \bar \beta })_{4\times
4}\otimes (A^{ab})_{3\times 3}$, in which the metric tensors of spinor part
and colour part become separable. This separable form gives an understanding
to colour confinement in the last paragraph. The understanding however, will
not be abated even if the two parts of metric tensor get entangled, since
anyway the condition $\mid A_{\bar \alpha \beta }^{ab}\mid =0$ makes the
rank of matrix $(A_{\bar \alpha \beta }^{ab})_{12\times 12}$ decrease at
least by {\bf one},{\bf \ which} belongs to either the spinor or colour
space. Therefore in what follows we take into account the general matrix
form with two parts entangled together. The following approximation of $A_{%
\bar \alpha \beta }^{ab}$ can be of an entangled form

\begin{equation}
(A_{\bar \alpha \beta }^{ab})=\gamma ^0\otimes I_{3\times 3}+A_\mu
\gamma ^0\gamma ^\mu \otimes \vec A_a\lambda ^a\text{ .}
\tag{11.14a}
\end{equation}
The use of Eq. (11.14a) can free us from writing out the explicit space-time
indices or colour indices in some special problems. For instance, in what
follows we can only write $\vec A$ to imply the inclusion of colour part, as
$\vec A=\vec A_a\lambda ^a$,

\begin{equation}
(A_{\bar \alpha \beta }^{ab})=\gamma ^0+A_\mu \gamma ^0\gamma ^\mu
\text{.} \tag{11.14b}
\end{equation}

For a perturbative interaction, it is impossible for scalar potential and
vector potential to satisfy the relation $\mid A_{\bar \alpha \beta }\mid
=1-A_0^2+\vec A^2=0\,$with $\vec A^2>0$. To avoid the coincidence with
perturbative interaction, for colour interaction we prefer the condition $%
\vec A^2>0$ rather than $\vec A\sim 0$, $A_0\sim 1$ to give the solution of $%
A_0^2=1+\vec A^2$. So we turn to the case $\vec A^2>0$ and $A_0^2>1$. We
recognize that this case doesn't rule out the possibility that while keeping
the difference $A_0^2-\vec A^2$ equal to $1$, we can at the same time make
both $\vec A\ $and $A_0$ continuously decrease until $A_0$ approaches the
{\it perturbative} situation $A_0\sim 1$with $\vec A^2\sim \varepsilon $%
(infinitesimal value). Therefore $\vec A^2>0$ seems also coincident with
{\it perturbative} situation, then we have to consider $\vec A^2<0$. Without
affecting the main conclusion, let's assume that the limiting situation $%
A_0^2=0$ holds. Under this situation, we have $\vec A=\pm i\vec A_W$ ($\vec A%
_W$ is the corresponding real vector).

Now let's take into account the explicit form $\vec A=\vec A_a\lambda ^a$.
Since every component of $\vec A$ is a matrix, for example $A_3=$ $%
A_a^3\lambda ^a$, the value of any component should be one of the
eigenvalues (of the corresponding matrix) or their combination ( the
coefficients $A_a^3$, $a=1,\cdots ,8$, are real) after tracing out the
degree of freedom of colour interaction by, e.g.
\begin{equation}
\left(
\begin{tabular}{lll}
$q_f^A$ & $q_f^B$ & $q_f^C$%
\end{tabular}
\right) A_a^3\lambda ^a\left(
\begin{array}{c}
q_i^a \\
q_i^b \\
q_i^c
\end{array}
\right) \text{ ,}  \tag{11.15}
\end{equation}
where the subscript $i$ and $f$ represent initial and final states
respectively. It can be asserted all values of components are real because
the generators $\lambda ^a$ ($a=1,\cdots ,8$) of group $SU(3)$ are hermitian
(we choose Gellmann matrices as the generators). Then it appears to be
inconsistent with the previous conclusion $\vec A=\pm i\vec A_W$ which is a
pure imaginary vector. One way to treat the inconsistency is to extend the $%
8 $--generators of $SU(3)$ to $9$--generators of $U(3)$
\begin{equation}
A_3=A_0^3\,i\,I_{3\times 3}+A_a^3\lambda ^a,\,a=1,\cdots ,8\text{
,} \tag{11.16}
\end{equation}
in which an additional pure imaginary generator $i\,I_{3\times 3}$ is
introduced, we denote it as $\lambda ^0$. We find $\lambda ^0$, with $%
\lambda ^0(\lambda ^0)^{\dagger }=1$, is not a Hermitian matrix, so that its
eigenvalues are not real. Now we conclude that when the confinement happens
only the first coefficient $\vec A_0$ is nontrivial, it yields
\begin{equation}
\vec A=i(A_0^1,\,A_0^2,\,A_0^3)\text{ .}  \tag{11.17}
\end{equation}
On condition that $A_0^2\neq 0$, the same extension of ''$8$--generators to $%
9$--generators'' would be necessary, and $\vec A$ is a complex number
accordingly.

The above extension of generators is reasonable because if a space is
curved, then some extra dimensions begin to get involved. For instance,
after a plane is curved, one must choose three-dimension space and the
corresponding transformation group in it to describe the curving. We choose $%
U(3)$ space as the curved quark space in order to satisfy the confining
condition.

In summary, in this subsection the equation $\mid A_{\bar \alpha \beta
}^{ab}\mid =0$ leads us to two explanations of the colour confinement:
First, it gives the decrease of the rank of colour matrix, which is a
general conclusion independent of separable--approximation. Second, it
induces the curving of colour space, which is a conclusion dependent on the
approximation (11.14) and (9.10).

\section{Summary and Discussion}

{\it Relationship with conventional Quantum Field Theory (CQFT)} The
geometry model presented in this paper can also be viewed as a theory for
elemental quantum fields, but it lacks many characteristics of CQFT. As for
symmetries, the transformations responsible for the groups $GL(n,\not C)$ or
$U(n,\not C)$ are used in this paper. Whereas the concept of gauge
transformation, under which the invariance of Lagrangian or Action is
respected in CQFT, is assumed to be irrelevant here. So the conservation
laws due to invariant Lagrangian are not appropriate to discuss in this
frame. However, from purely geometrical angle, we respect the invariance of
geometric quantities such as metric and curvature etc. under the
transformation of structure group. Generally speaking, in a nonlocal theory
any conservation law holds only when some integrations are carried out over
certain spaces, and thus the square root of Jaccobian should appear in the
forms of conservation law, as that in GR and implied in Sections 8\symbol{126%
}10.

Furthermore, generally in CQFT the Lagrangian depends on the fields and
their first derivatives only. But in differential geometric, it allows for
up to two derivatives of fields, which can be noticed in the quadratic form
of motion equation for fermions and field equation for bosons.

In view of the distinguished success of CQFT in describing the
purterbative interactions, the results from these sectors are
respected. First, the Dirac equation and Klein-Gordon equation in
QED, the asymptotically freedom in QCD etc. are employed to
compare with the results of our model. In that respect, the
quadratic forms of Dirac equation and Klein-Gordon equation are
regained under some approximations. Second, the interaction
vertices of QED and QCD are used to design appropriate
approximations of metric tensor used in the second half of the
paper. However, the technique of renormalization is assumed
unnecessary in this theory. Because the inclusion of the nonlocal
characteristics of quantum wave in this theory would automatically
make the value of particles' momentum within a limited range.

{\it Relationship with General Relativity (GR) }As stated in the
introduction, the formalism developed in this work is similar to
that of GR. The difference is that we have generalized the space
from real to complex. The important characteristic of our model is
that the base manifold is complex. If the space-time is initially
used as the base manifold, it is impossible to clarify physical
meanings of dynamics or to get the results of this paper.
Additionally, the manifolds under concern are not necessarily
Riemann manifold, i.e. the metric tensor is not necessarily symmetric, $%
A(X,Y)=A(Y,X)$.

In GR, only one type of curving--Riemann curving (hyperbolic
curving)--appears, but in our theory two curvings are present--both
hyperbolic and elliptic curvings. They are represented by two different
structure groups $GL(n,\not C)$ and $U(n,\not C)$, which correspond to two
differential geometries. The geometry with structure group $U(n,\not C)$ is
relevant to Riemann manifold [e.g. $1/2(A(X,Y)+A(Y,X))$ gives Riemann
metric], but $GL(n,\not C)$ not. In non-Abelian case, the metric tensors
from two geometries may be entangled as shown in (11.14).

{\it The Approximations Used in This Paper} All the meaningful results are
dependent on the four approximations in Eqs. (9.10), (10.13), (11.10) and
(11.14). The credibility of the results relies on the reasonability of the
approximations. The approximations follow from our understanding of the
interaction vertex in CQFT. The appreciated aspect of these approximations
is the restoration of the terms of quadratic Dirac equation and Klein-Gordon
equation by the aids of the replacement $d\rightarrow \gamma _\mu \partial
^\mu $. On the other hand, the appearance of mass term for bosons and the
understanding to quark confinement also follow from above approximations and
the replacement.

From complex space to real space we use $d\rightarrow \gamma _\mu \partial
^\mu $ as a projection, we have to do so to get some physical observables in
space-time. In doing so some information is inevitably lost, but it cannot
be avoided--the observed space is complex, the observing space is real.
Strictly speaking, the replacement $d\rightarrow \gamma _\mu \partial ^\mu $
is not an approximation, but rather a technique following the approximation $%
(A^{\alpha \bar \rho })_{4\times 4}=\gamma _0-\gamma _0\gamma ^\mu A_\mu $.
It helps us gain the form of motion equation for fermions, though with quite
a few additional integrations involved. To resolve that intricate motion
equation possibly requires us to convert it back to the Dirac-equation form
by using some special techniques.

{\it Remarks on applicability of the theory} Although the model is designed
for a low-energy nonperturbative interaction, at present almost all the
approximations imposed on the model have perturbative forms and
originations. And thus all the results here rely on both the theory and the
approximations. We have noticed some results, such as the additional terms
in motion equation and the meaning of $\mid A_{\bar \alpha \beta }^{ab}\mid
=0$ etc., do not completely fit to the perturbative cases, which is expected
to be the nonperturbative signals of our model. So after the model is
reduced to get the main essences of CQFT, we hope it can be perfectly useful
for nonperturbative interaction [e. g. in phenomenological study of the
light--quark excited states.] or, long--wavelength low--energy region of
perturbative interaction [e. g. strong correlations of fermions in ultra low
temperature.], as anticipated initially.

{\it Future Development of the theory} First, we should search for
a smaller group than $GL(4,C)$ to accurately describe the
hyperbolic curving (Lorentz violation of spinors' transformation),
as stated in Sec. 7. The search progress may involve some new
physics, not a merely mathematical problem. Second, we should
present a more reliable way to give the mass term in fermion
motion equation in Sec. 8. C, some numerical calculation may be
involved. Third, we hope the theory applicable to weak interaction
too. Relevantly, the final understanding of mass problem relies
undoubtedly on the issues of weak interaction.

\section*{Acknowledgments}

A lot of thanks to Prof. W. T. Geng for his constructive
suggestions in preparing this manuscript.

\section{Appendix. A Evaluating the integral of Eq. (8.10)}

In this appendix we give the detailed steps and methods of evaluating the
first integration in Eq. (8.10).

\begin{eqnarray}
&&\ \ \ \int (1-\gamma _\mu A^\mu )\gamma _\sigma \partial ^\sigma (1+\gamma
_\nu A^\nu )\gamma _\rho \gamma _\lambda \partial ^\rho { \psi \,}%
dx^\lambda \text{ }  \nonumber \\
\ &=&\int \gamma _\sigma \partial ^\sigma (\gamma _\nu A^\nu )\gamma _\rho
\gamma _\lambda \partial ^\rho { \psi \,}dx^\lambda  \nonumber \\
&&\ \ \ -\int \gamma _\mu A^\mu \gamma _\sigma \partial ^\sigma
(\gamma _\nu A^\nu )\gamma _\rho \gamma _\lambda \partial ^\rho {
\psi \,}dx^\lambda \text{,}   {(8.10)} \nonumber
\end{eqnarray}
here ${ \tilde \psi =}\{{ \psi }^1,{ \psi }^2,{ \psi }^3,{ %
\psi }^4\}$ is a four-component vector, $A^{\alpha \bar \rho }\,$ and $A_{%
\bar \rho \gamma }$ are all in their matrix forms, $1-\gamma _\mu A^\mu $
and $1+\gamma _\nu A^\nu $. We will mainly deal with the first integration
in (8.10), since the second term includes two $A^\mu $ factors we
temporarily omit it as second-order perturbation.

\begin{eqnarray}
&&\int \gamma _\sigma \partial ^\sigma (\gamma _\nu A^\nu )\gamma _\rho
\gamma _\lambda \partial ^\rho { \psi \,}dx^\lambda  \nonumber \\
&=&\int \gamma _\sigma \gamma _\nu (\partial ^\sigma A^\nu )\gamma _\rho
\gamma _\lambda \partial ^\rho { \psi \,}dx^\lambda  \nonumber \\
&=&\int (g_{\sigma \nu }+\underline{\gamma _\sigma \gamma _\nu })(\partial
^\sigma A^\nu )\gamma _\rho \gamma _\lambda \partial ^\rho { \psi \,}%
dx^\lambda  \nonumber \\
&=&\int \partial _\nu A^\nu \gamma _\rho \gamma _\lambda \partial ^\rho { %
\psi \,}dx^\lambda -\int \partial ^\sigma (\gamma _\nu A^\nu
)\gamma _\sigma \gamma _\rho \gamma _\lambda \partial ^\rho { \psi
\,}dx^\lambda      {(A. 1)} \nonumber
\end{eqnarray}
henceforth in this Appendix we use $\underline{\gamma _\sigma \gamma _\nu }$
to express the product $\gamma _\sigma \gamma _\nu $ when $\sigma \neq \nu $%
. We evaluate the above two terms in (A.1) separately. The first term
\begin{eqnarray}
&&\int \partial _\nu A^\nu \gamma _\rho \gamma _\lambda \partial ^\rho { %
\psi \,}dx^\lambda  \nonumber \\
&=&\int \partial _\nu A^\nu (g_{\rho \lambda }+\underline{\gamma _\rho
\gamma _\lambda })\partial ^\rho { \psi \,}dx^\lambda  \nonumber \\
&=&\int \partial _\nu A^\nu \partial _\lambda { \psi \,}dx^\lambda
+\frac 12\gamma _\rho \gamma _\lambda \int \partial _\nu A^\nu
(dx^\lambda \partial
^\rho -dx^\rho \partial ^\lambda ){ \psi \,}  \nonumber \\
&=&\partial _\nu A^\nu { \psi \mid }_{-\infty }^t-\int { \psi (}%
\partial _\lambda \partial _\nu A^\nu ){ \,}dx^\lambda +\frac 12\gamma
_\rho \gamma _\lambda \int \partial _\nu A^\nu (dx^\lambda
\partial ^\rho -dx^\rho \partial ^\lambda ){ \psi \,\,}\text{
.}   {(A. 2)} \nonumber
\end{eqnarray}
To evaluate the second term in (A.1), the following relation is useful:
\begin{equation}
\gamma _\sigma \gamma _\rho \gamma _\lambda =\{
\begin{array}{c}
-i\,\epsilon _{\sigma \rho \lambda \mu }\gamma _5\gamma ^\mu \text{ , if }%
\sigma \text{, }\rho \text{, }\lambda \text{ unequal to each other,} \\
g_{\sigma \rho }\gamma _\lambda +g_{\rho \lambda }\gamma _\sigma -g_{\sigma
\lambda }\gamma _\rho \text{, if at least two of }\sigma \text{, }\rho \text{%
, }\lambda \text{ equal to each other.}
\end{array}
\tag{A. 3}
\end{equation}
Substituting this relation to second term in (A.1) yields,

\begin{eqnarray}
&&\int \partial ^\sigma (\gamma _\nu A^\nu )\gamma _\sigma \gamma _\rho
\gamma _\lambda \partial ^\rho { \psi \,}dx^\lambda  \nonumber \\
&=&-i\,\epsilon _{\sigma \rho \lambda \mu }\gamma _5\gamma ^\mu \int
\partial ^\sigma (\gamma _\nu A^\nu )\partial ^\rho { \psi \,}dx^\lambda +
\nonumber \\
&&\int \partial ^\sigma (\gamma _\nu A^\nu )(g_{\sigma \rho
}\gamma _\lambda +g_{\rho \lambda }\gamma _\sigma -g_{\sigma
\lambda }\gamma _\rho )\partial ^\rho { \psi \,}dx^\lambda  {(A.
4)} \nonumber
\end{eqnarray}
We will leave the first term of (A. 4) as it is and turn to cope with the
second term,
\begin{eqnarray}
&&\int \partial ^\sigma (\gamma _\nu A^\nu )(g_{\sigma \rho }\gamma _\lambda
+g_{\rho \lambda }\gamma _\sigma -g_{\sigma \lambda }\gamma _\rho )\partial
^\rho { \psi \,}dx^\lambda  \nonumber \\
&=&\int \partial ^\sigma A^\nu \gamma _\nu \gamma _\lambda
\partial _\sigma { \psi \,}dx^\lambda +\int \partial ^\sigma A^\nu
\gamma _\nu \gamma _\sigma \partial ^\lambda { \psi \,}dx_\lambda
+\int \partial ^\lambda A^\nu \gamma _\nu \gamma _\rho \partial
^\rho { \psi \,}dx_\lambda
\nonumber \\
&=&\int \partial ^\sigma A^\nu (g_{\nu \lambda }+\underline{\gamma
_\nu \gamma _\lambda })\partial _\sigma { \psi \,}dx^\lambda +\int
\partial
^\sigma A^\nu (g_{\nu \sigma }+\underline{\gamma _\nu \gamma _\sigma }%
)\partial ^\lambda { \psi \,}dx_\lambda  \nonumber \\
&&+\int \partial ^\lambda A^\nu (g_{\nu \rho }+\underline{\gamma _\nu \gamma
_\rho })\partial ^\rho { \psi \,}dx_\lambda  \nonumber \\
&=&\int \partial ^\sigma A^\nu \partial _\sigma { \psi \,}dx_\nu +\frac 12%
\gamma _\nu \gamma _\lambda \int (dx^\lambda \partial ^\sigma A^\nu -dx^\nu
\partial ^\sigma A^\lambda )\partial _\sigma { \psi \,}  \nonumber \\
&&+\int \partial _\nu A^\nu \partial ^\lambda { \psi \,}dx_\lambda
+\frac 12\gamma _\nu \gamma _\sigma \int F^{\sigma \nu }\partial
^\lambda { \psi
\,}dx_\lambda  \nonumber \\
&&+\int \partial ^\lambda A_\nu \partial ^\nu { \psi \,}dx_\lambda
-\frac 12\gamma _\nu \gamma _\rho \int (\partial ^\lambda A^\nu
\partial ^\rho -\partial ^\lambda A^\rho \partial ^\nu ){ \psi
\,}dx_\lambda   {(A. 5)} \nonumber
\end{eqnarray}
where $F^{\sigma \nu }=(\partial ^\sigma A^\nu -\partial ^\nu A^\sigma )$.
Then performing the integration by parts, it yields

\begin{eqnarray}
&&\int \partial ^\sigma (\gamma _\nu A^\nu )(g_{\sigma \rho }\gamma _\lambda
+g_{\rho \lambda }\gamma _\sigma -g_{\sigma \lambda }\gamma _\rho )\partial
^\rho { \psi \,}dx^\lambda  \nonumber \\
&=&\gamma _\nu \gamma _\lambda \int \partial ^\sigma A^\nu \partial _\sigma
{ \psi \,}dx^\lambda { +}  \nonumber \\
(\partial _\nu A^\nu ){ \psi } &{ \mid }_{-\infty }^t-\int { \psi (}%
\partial _\lambda \partial _\nu A^\nu ){ \,}dx^\lambda -\frac 12\gamma
_\nu \gamma _\sigma F^{\nu \sigma }{ \psi \mid }&_{-\infty }^t+
\nonumber
\\
&&\frac 12\gamma _\nu \gamma _\sigma \int { \psi (}\partial
_\lambda
F^{\nu \sigma }){ \,}dx^\lambda -  \nonumber \\
A^\nu \partial _\nu { \psi } &{ \mid }_{-\infty }^t+\int A^\nu { (}%
\partial _\lambda \partial _\nu { \psi }){ \,}dx^\lambda -\frac 12%
\gamma _\nu \gamma _\rho \tilde F^{\nu \rho }{ \psi \mid
}&_{-\infty }^t+
\nonumber \\
&&\frac 12\gamma _\nu \gamma _\rho \int \tilde F^{\nu \rho
}\partial _\lambda { \psi \,}dx^\lambda \text{ ,}   {(A. 6)}
\nonumber
\end{eqnarray}
where $\tilde F^{\mu \nu }=A^\mu \partial ^\nu -A^\nu \partial ^\mu $. In
the first term of the right hand we have converted the expansion $\int
\partial ^\sigma A^\nu \partial _\sigma { \psi \,}dx_\nu \,+\frac 12%
\gamma _\nu \gamma _\lambda \int (dx^\lambda \partial ^\sigma A^\nu -dx^\nu
\partial ^\sigma A^\lambda )\partial _\sigma { \psi }$ back to its
original form for there is no way integrating out the coordinates to simple
form. Now substituting Eq. (A.6) and Eq. (A.2) into Eq. (A.1) leads to

\begin{eqnarray}
&&\ \int \gamma _\sigma \partial ^\sigma (\gamma _\nu A^\nu )\gamma _\rho
\gamma _\lambda \partial ^\rho { \psi \,}dx^\lambda  \nonumber \\
\ &=&A_\nu \partial ^\nu { \psi +}\frac 12\gamma _\mu \gamma _\nu \tilde F%
^{\mu \nu }{ \psi +}\frac 12\gamma _\mu \gamma _\nu F^{\mu \nu }{ %
\psi }  \nonumber \\
&&\ +\frac 12\gamma _\rho \gamma _\lambda \int (\partial _\nu
A^\nu )(dx^\lambda \partial ^\rho -dx^\rho \partial ^\lambda ){
\psi \,}
\nonumber \\
&&\ +i\epsilon _{\sigma \rho \lambda \mu }\gamma _5\gamma ^\mu
\int \partial ^\sigma (\gamma _\nu A^\nu )\partial ^\rho { \psi
\,}dx^\lambda -\int
A_\nu \partial _\lambda \partial ^\nu { \psi \,}dx^\lambda -\frac 12%
\gamma _\mu \gamma _\nu \int \tilde F^{\mu \nu }\partial _\lambda
{ \psi
\,}dx^\lambda  \nonumber \\
&&\ -\frac 12\gamma _\mu \gamma _\nu \int \partial _\lambda F^{\mu \nu }{ %
\psi \,}dx^\lambda -\gamma _\mu \gamma _\nu \int \partial ^\sigma A^\mu
\partial _\sigma { \psi \,}dx^\nu \text{ .}   {(A. 7)} \nonumber
\end{eqnarray}

\section*{Figure Captions}

Fig. 1: To describe how one fermion observes (interacts with)
another fermion, we employ the formalism of General Relativity by
generalizing its space from real to complex. In gragh, only three
axes are displayed, in fact we work in four-dimension space.

Fig. 2: Patterns of physical region in three-dimension space,
respectively for electrons (A) and quarks (B). Likewise the
singularity and physical region for complex space are defined in
Sec. 11 by adding constraints to interaction potential.

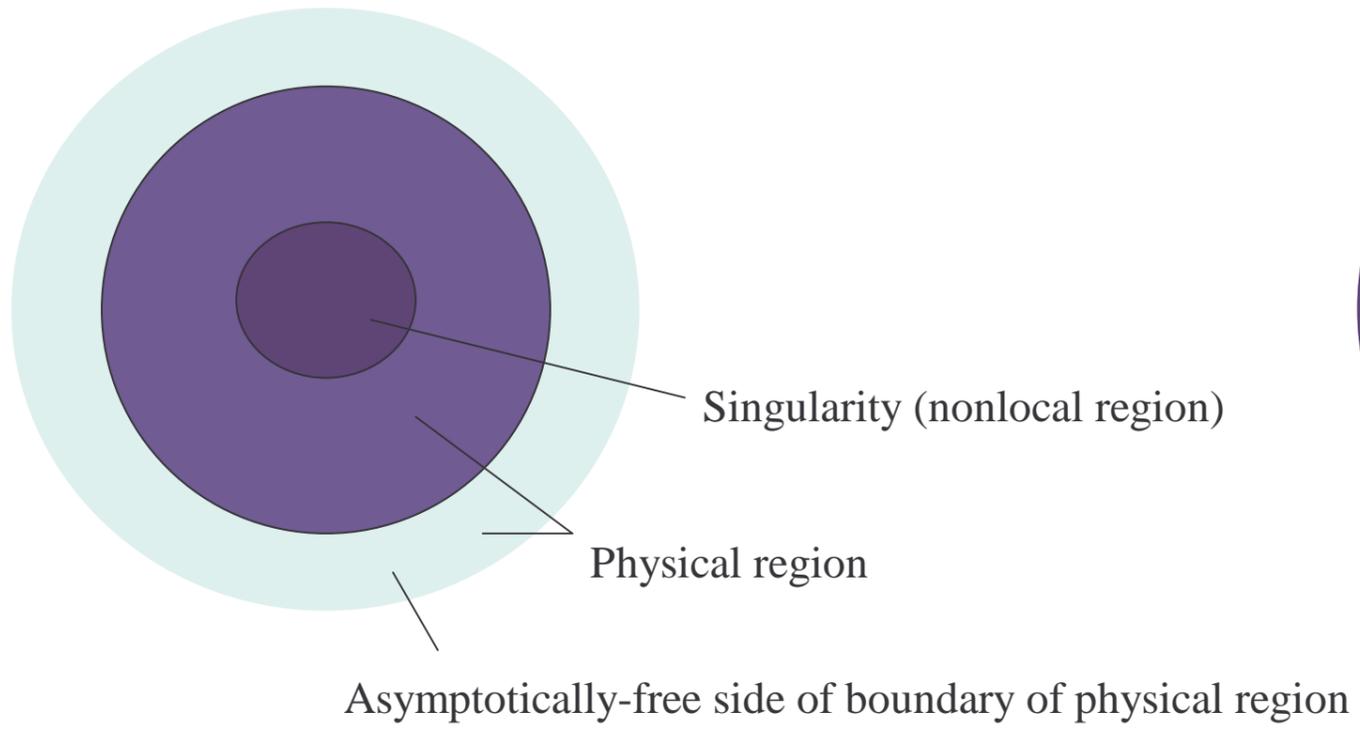
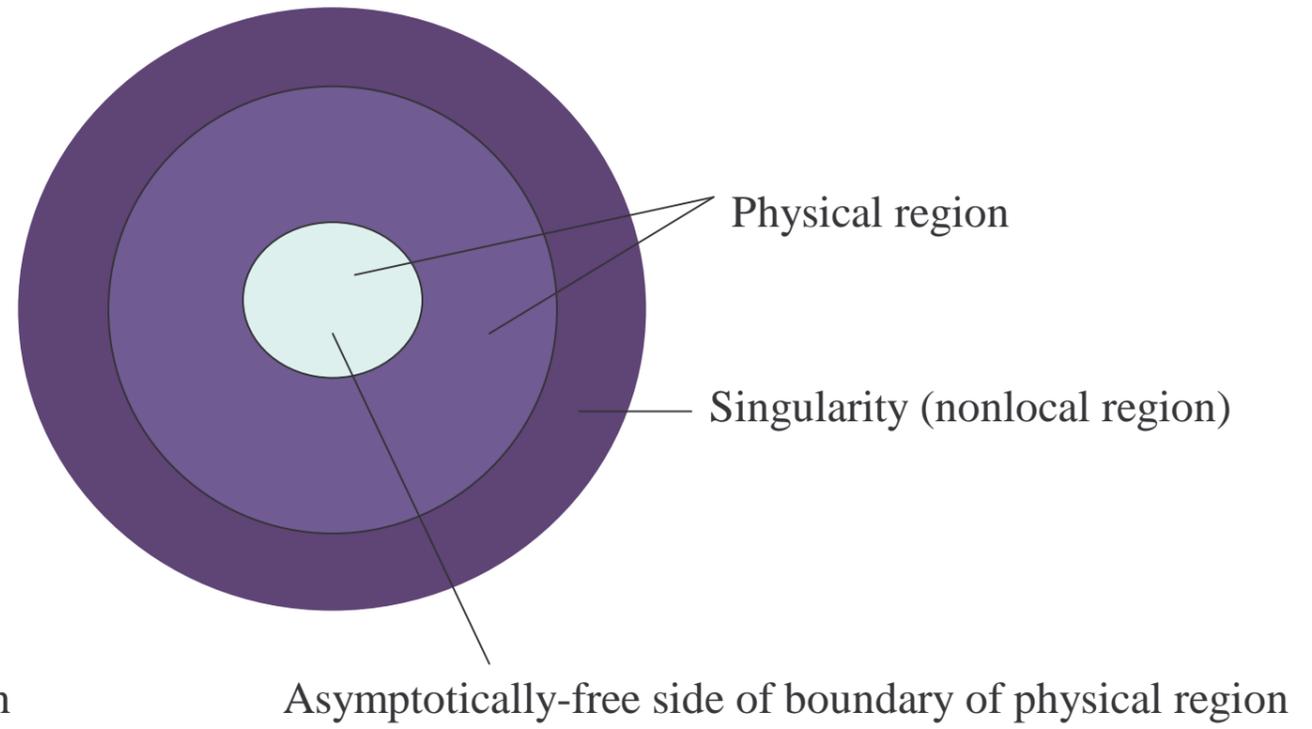
\end{document}